\documentclass[apj]{emulateapj}
\usepackage{mathptmx}

\def\d{\mathrm{d}}

\begin{document}

\title{On the connection between metal absorbers and quasar nebulae}

\author{Doron Chelouche\altaffilmark{1,2}, 
Brice M\'enard\altaffilmark{3,1}, 
David V. Bowen\altaffilmark{4},
and Orly Gnat\altaffilmark{5}}

\altaffiltext{1} {School of Natural Sciences, Institute for Advanced
Study, Einstein Drive, Princeton 08540, USA; doron@ias.edu}

\altaffiltext{2} {Chandra Fellow}

\altaffiltext{3} {Canadian Institute for Theoretical Astrophysics,
University of Toronto, 60 Saint George Street, Toronto, ON M5S 3H8,
Canada}

\altaffiltext{4} {Department of Astrophysical Sciences, Princeton
University, Princeton, NJ 08544}

\altaffiltext{5} {School of Physics and Astronomy and the Wise
Observatory; Tel-Aviv University, Tel-Aviv, Israel} 

\shortauthors{D. Chelouche et al.}

\shorttitle{Metal Absorbers \& Quasar Nebulae}

\begin{abstract}

We establish a simple model for the distribution of cold gas around
$L^\star$ galaxies using a large set of observational constraints on
the properties of strong \ion{Mg}{2}\ absorber systems.  Our analysis suggests
that the halos of $L^\star$ galaxies are filled with cool gaseous
clouds having sizes of order 1\,kpc and densities of $\sim 10^{-2}$
cm$^{-3}$.  We then investigate the physical effects of cloud
irradiation by a quasar and study the resulting spectral
signatures. We show that quasar activity gives rise to (i) extended
narrow-line emission on $\sim100$ kpc scales and (ii) an anisotropy in
the properties of the absorbing gas arising from the
geometry of the quasar radiation field.  Provided that quasars reside in
halos several times more massive than those of $L^\star$ galaxies, our
model predictions appear to be in agreement with observations of
narrow emission-line nebulae around quasars and the recent detections
of $\sim100$\,kpc cold gaseous envelopes around those objects,
suggesting a common origin for these phenomena.  We discuss the
implications of our results for understanding absorption systems,
probing quasar environments at high redshifts, and testing the quasar
unification scheme.

\end{abstract}


\section{Introduction}

Observational data on quasar absorption lines indicates that $L^\star$
galaxies are surrounded by cool ($\sim 10^4$\,K) gas condensations
ranging up to radii of $\sim
100\,h_{70}^{-1}$\,kpc\footnote{Throughout the paper we use
$H_0=70\,h_{70}~{\rm km~s^{-1} Mpc^{-1}}$, $\Omega_M=0.7$, and
$\Omega_\Lambda=0.3$.} (Bergeron et al. 1987, Steidel et al. 1995,
Churchill et al. 2005, Zibetti et al. 2006). Similar distributions
have recently been found around quasars (Bowen et al. 2006, 2007).  In
addition, several teams have discovered that bright quasars are
surrounded by giant line-emitting nebulae observed at transitions such
as Ly$\alpha$ (e.g., Wampler et al. 1975, Heckman et al. 1991a,b,
Christensen et al. 2006).  In this paper we investigate whether the
two phenomena could originate from the same type of gas clouds around
galaxies thereby connecting gas absorption and line emission in galaxy
and quasar halos.
 
Quasar spectra have historically been a main source of information
about large scale matter distribution in the universe. Such
information has been obtained by studying absorption lines induced by
intervening material between us and the quasar. In the optical
wavelength range, absorption lines by metal-enriched gas at $z
\lesssim 2$ originate from low to moderate-ionization species and the easiest
feature to detect is usually the \ion{Mg}{2}\ doublet ($\lambda\lambda
2796.35,2803.53$). It is the first resonance transition of an abundant
element with a large oscillator strength to enter the optical range
and has therefore been the most commonly used tracer of cool
gas. Several studies have shown that \ion{Mg}{2}\ arises in gas
spanning more than five orders of magnitude in neutral hydrogen column
density, from $N_{\rm HI} \simeq 10^{17}-10^{22}~{\rm cm^{-2}}$
\citep{bergeron_stasinska_86, steidel_sargent_92,churchill+00}.
Following the suggestion by Bahcall
\& Spitzer (1969) that quasar absorption lines might arise in halos around galaxies, an association between absorbers and galaxies has been
reported, starting from small samples of a few to $\sim50$ galaxies
(\citealp{bergeron_86,cristiani_87,bergeron_boisse_91,1997ApJ...480..568S};
Churchill et al. 2005) to statistical measurements involving thousands
of objects (Zibetti et al. 2005 \& 2006, M{\'e}nard et al. 2007).

While a connection between galaxies and metal absorbers has been
firmly established, the origin of the absorbing gas remains unclear.
Whether these absorption line systems trace gas being accreted by a
galaxy or outflowing from it is still a matter of debate. Among the
possible origins for absorption line systems in galaxy halos are condensations of
cool clouds from hot and dilute halo gas via thermal instability
(e.g., Mo \& Miralda-Escude 1996, Maller \& Bullock 2004), cool gas
that is bound to dark matter sub-halos embedded within the main halo
(e.g., Sternberg, McKee, \& Wolfire 2002 and references therein), and
starburst driven winds from galaxies (e.g., Oppenheimer \& Dav{\'e}
2006, Prochter et al. 2006). Additional explanations include (warped) galaxy disks (Bowen et
al. 1995, Prochaska \& Wolfe 1998) as well as low surface brightness
companion objects (York et al. 1986, Petitjean \& Bergeron 1994).

Some galaxies contain active nuclei and it is thought that all quasars
are harbored in galaxies. The properties of quasar hosts are not well
characterized but it is believed that bright quasars are usually
harbored in rather massive, star-forming galaxies ranging from $\sim$
1 to $\sim 30 L^\star$ (Jahnke \& Wisotzki 2003).  Little is known
about the environments of quasars, with recent studies suggesting that
brightest objects live in somewhat over-dense environments (Serber et
al. 2006). Like galaxies, the immediate environment of quasars can
also be probed in absorption: using a survey of projected quasar pairs
from the {\it Sloan Digital Sky Survey} (SDSS) our team has recently
showed that quasars, like galaxies, are surrounded by Mg\,II gas on
scales ranging up to 100\,$h_{70}^{-1}$\,kpc (Bowen et
al. 2006). Using a larger sample of quasar pairs, we have now found
that the transverse gas distribution around quasars, as traced by
Mg\,II, suggests a covering factor of order unity up to $\sim
100\,h_{70}^{-1}$ kpc (Bowen et al. 2007). 

Gas which is associated with the quasar environment is also commonly
observed (via resonance line absorption such as \ion{C}{4}\,$\lambda
1548$) in our line-of-sight toward quasars and is identified by its
proximity (within $\sim10^3~{\rm km~s^{-1}}$) to the quasar
redshift. Nevertheless, it is generally hard to establish whether such
gas lies in the vicinity of the black hole (on sub-pc scales) or
whether it originates in its host galaxy or its surrounding halo
(see Crenshaw et al. 2003 for a review).

Deep imaging and spectroscopic observations of quasars reveal that a
significant fraction of them have surrounding emission-line
nebulosities extending out to $\sim 100\,h_{70}^{-1}$\,kpc. Such
nebulosities are often detected in Ly$\alpha$ emission (e.g.,
Heckman et al. 1991, Christensen et al. 2006) or the
[\ion{O}{3}]\,$\lambda 5007$ line (Stockton \& MacKenty 1987, Fu \&
Stockton 2006) with recent surveys suggesting that radio-loud quasars
(RLQ) may have brighter Ly$\alpha$ nebulosities (Christensen et
al. 2006).  Little is known about the physical nature of these nebulae
and various scenarios pertaining to both gas accretion onto the quasar
as well as gas ejection from it have been put forward as possible
explanations (e.g., Heckman et al. 1991, Haiman \& Rees 2001).

Motivated by these recent findings, we investigate in this work
whether the emitting gas seen around quasars is consistent with being
the same material as that observed around galaxies (and quasars) in
absorption.  The paper is organized as follows: In \S 2 we first
summarize some observational constraints on the properties of cool gas
around galaxies used to calibrate our model. A geometrical model is
then constructed and its physics is discussed.  We then (\S 3) use
this model to study the effects on the absorption and emission
properties of halo gas when exposed to a quasar radiation field. In
addition, we show that such a model can naturally explain the
distribution of cool gas around quasars and the existence of
large-scale Ly$\alpha$ emitting nebulae. We discuss the implications
of our results for galaxy formation, baryon fraction in the Universe,
and quasar physics in \S 4. Summary follows in \S 5.

\section{The distribution of cool gas\\ around $L^\star$ galaxies}
\label{section_model}

\subsection{Observational constraints from \ion{Mg}{2}\ absorbers}

Observational data on \ion{Mg}{2}\ absorption lines have provided  a
number of constraints on the spatial distribution and characteristics
of the cool gas around galaxies. Here we list the most relevant
results that can be used to model the \ion{Mg}{2}\ distribution in the halo of
an $L^\star$ galaxy:

\textbf{1.)~} Strong \ion{Mg}{2}\ systems, defined with a rest equivalent
width $W_0>0.3$\AA,  inhabit the halos of $\sim L^\star$
galaxies, as was found by Steidel et al. (1997). These authors found that Mg\,II
selected galaxy luminosities lie in the range $0.3-5\,L^\star$ with a
mean $\langle L \rangle\simeq 0.8\,L^\star$.  Their results also
suggested a unit covering factor up to $\sim$50\,$h_{70}^{-1}$\,kpc
and the absence of strong \ion{Mg}{2}\ absorption on larger scales.  More
recent studies (Zibetti et al. 2005 \& 2006, Churchill et al. 2005)
indicate that strong Mg\,II absorption can actually be detected on
scales reaching more than 100\,$h_{70}^{-1}$\,kpc but that the
covering factor falls off in this range. We note that the study done
by Zibetti et al. (2006) implies little evolution in the properties of
absorbing galaxies up to $z\sim 1$.

\textbf{2.)~} Strong \ion{Mg}{2}\ absorbers usually show saturated absorption
lines (as implied by the doublet ratio).  High-resolution
spectroscopy reveals that strong systems usually break into several
kinematically distinct components (e.g., Churchill \& Vogt 2001). When resolved,
the widths of these components are consistent with gas at a temperature of $\sim
10^4$\,K (e.g., Lanzetta \& Bowen 1990). It has been established that
the rest equivalent width of strong systems is roughly proportional to the
number of distinct kinematical components.  Bergeron \& Petitjean
(1990) found the relation between total rest equivalent width and
number of clouds to be $W_0({\rm {Mg}\,{II}}) \simeq 0.3\times N_{\rm
c}$\,\AA\ and Churchill (1997) reported $W_0({\rm {Mg}\,{II}}) \simeq
0.07\times N_c$\,\AA\ where $N_{\rm c}$ is the number of clouds
along the line-of-sight. If the individual clouds observed in these
systems are relatively similar, then these results suggest that, even
if the rest equivalent width of \ion{Mg}{2}\ systems is mostly driven by a
velocity dispersion, it can be used as a proxy for the number of
clouds along the line of sight.

\textbf{3.)~} Constraints on the sizes of \ion{Mg}{2}\ systems have been
obtained by studying the spectra of strongly lensed quasars. 
Limits on the size of individual absorption components are
obtained by comparing absorption features across different lines of
sight. Typical values span the range $0.1-10$
kpc for \ion{Mg}{2}\ clouds (e.g., Rauch et al. 2002, Ellison et al. 2004)
and perhaps somewhat larger sizes for systems detected by high
ionization lines such as \ion{C}{4}\,$\lambda\lambda 1548,1550$.  At
present this is the only direct means for estimating the size of cool
gaseous clouds in galaxy halos (for an indirect method relying on
photoionization modeling see e.g., Ding et al. 2003).

\textbf{4.)~} Finally, HI measurements show that strong \ion{Mg}{2}\ systems with
$W_0\sim 0.5$\AA\ give rise to HI column densities in the range
$10^{18-20}$ cm$^{-2}$ while stronger systems with $W_0\sim 2$\AA\
have $N_{\rm HI}\sim 10^{19-21}$ cm$^{-2}$ (Rao et al. 2006).
Emission-based results from 21 cm surveys at $z=0$ (Zwaan et al. 2005)
indicate that the contribution of galaxy \emph{disks} to the observed
HI column density distribution can be substantial within $\sim$20 kpc
for certain orientations.  (See also Bowen et al. 1995 and Prochaska
\& Wolfe 1997). However, on larger scales, the contribution of halo
gas clouds is in general expected to dominate the HI column density
distribution.  The lack of correlation between \ion{Mg}{2}\ absorption
properties and galaxy size or inclination further supports this
assertion (Steidel et al. 2002; see also Kacprzak et al. 2007).

\subsection{Modeling the spatial distribution}

\begin{figure*}
\begin{center}
\plottwo{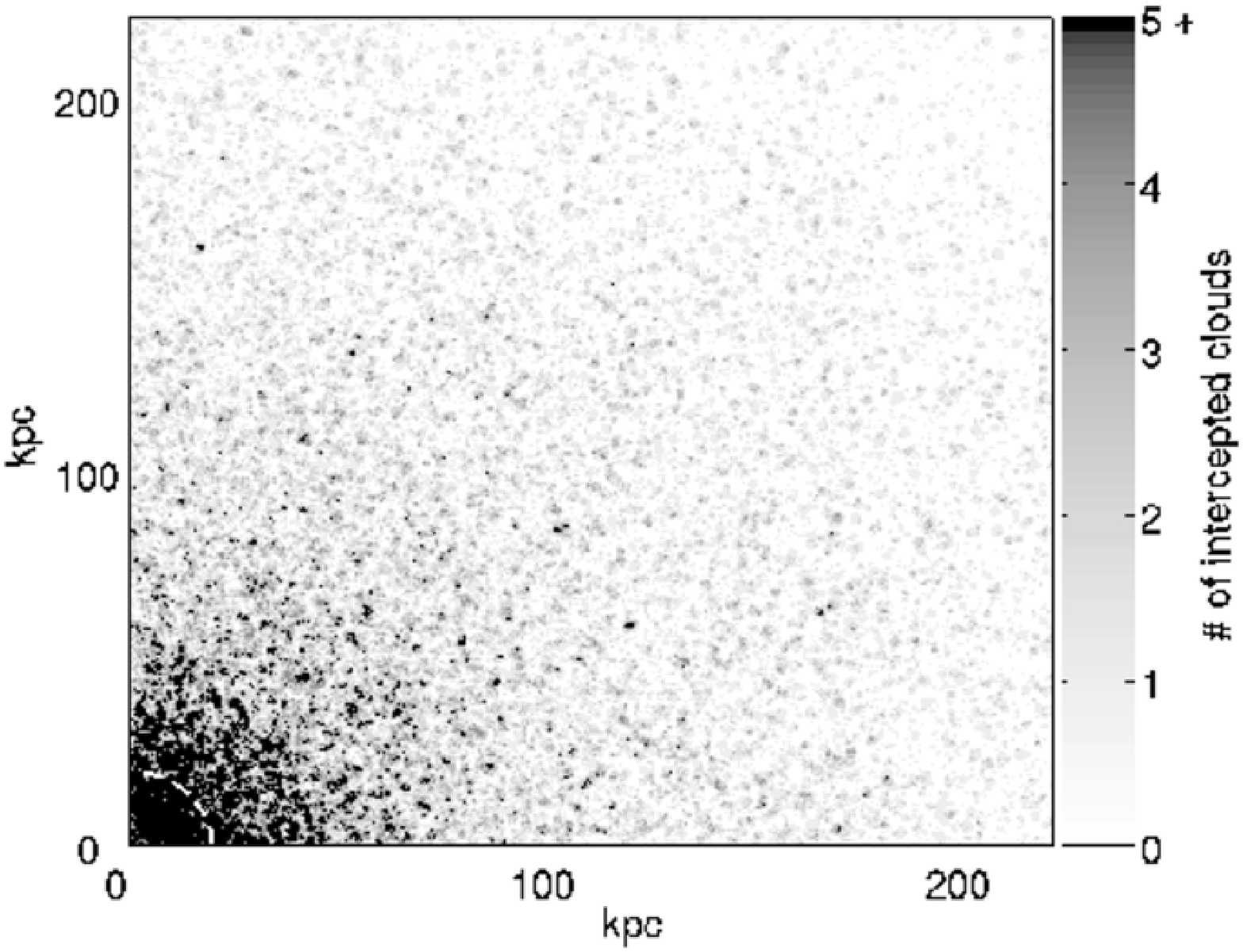}{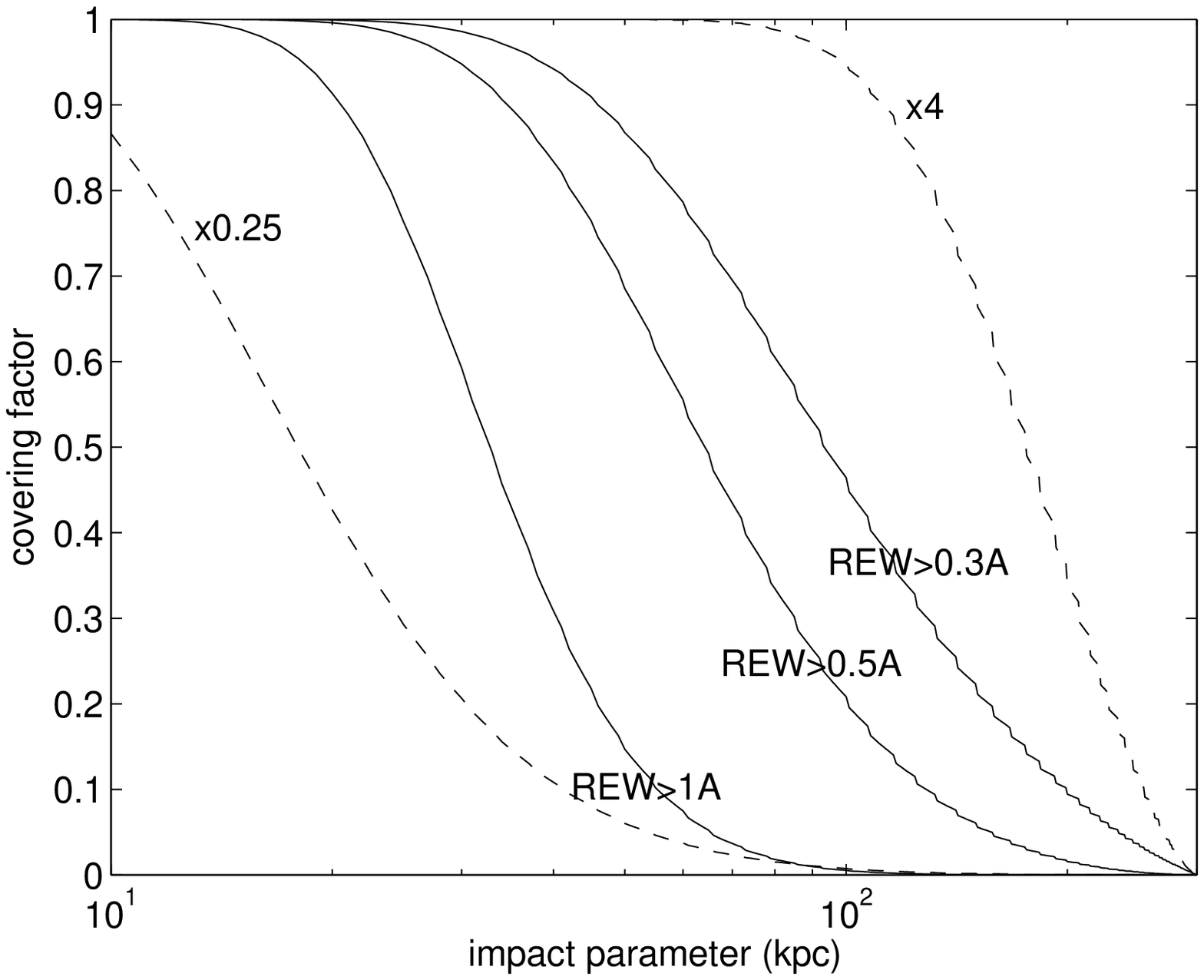}
\caption{The number-count distribution of Mg\,II clouds along
different lines-of-sight through a halo of an $L^\star$ galaxy (see
text). Note the increase in the number of intercepted cloud with
decreasing impact parameter. Note also that clouds are distributed up
to the virial radius though the probability for intercepting them on
large scales is small. \emph{Right:} The covering factor for strong
Mg\,II absorption around galaxies. The different solid lines
correspond to model predictions of the covering factor for different
$W_0$ limits (denoted to the right of each curve) around an $L^\star$
galaxy. As shown, the covering factor for strong, $W_0>0.5$\AA\
absorption is below 50\% for impact parameters greater than
50\,kpc. Conversely, weak absorption with $W_0<0.3$\AA\ is common for
impact parameters greater than 70\,kpc.  The dashed lines are
model predictions for galaxy halo models with cool gas mass 0.25 and 4
times that of an $L^\star$ galaxy and for $W_0>0.5$.}
\end{center}
\label{fig_model_realization}
\end{figure*}

We now introduce a simple phenomenological model aimed at reproducing
the main aspects of the observational constraints presented above,
namely the spatial extent of the gas, the cross-section for strong
\ion{Mg}{2}\ absorption and the distribution of hydrogen column densities
observed around $\sim L^\star$ galaxies.  We emphasize that, in this
study, we do \emph{not} attempt to investigate the physical origin of
the gas, i.e. whether it is bound to satellite dark matter halos,
infalling or being ejected out of the galaxy via winds.  We consider a
case in which the total mass density (baryonic and dark matter) for
such a system is described by an isothermal profile:
\begin{equation}
\rho(r)=\frac{\sigma^2}{2\pi G r^2}\,, ~~~~~r\le
r_v=\frac{\sigma}{\sqrt{50}\,H_0}
\end{equation}
where $\sigma$ is the velocity dispersion, $r_v$ the virial radius,
and $\rho$ is the density. The mass within a given radius $r$ is then
given by
\begin{equation}
M(<r)=\frac{2\,\sigma^2\,r}{G}
\end{equation}
and by definition, the density enclosed within the virial radius is
200 times the critical density.  According to Fukugita \& Peebles
(2006), the one-dimensional mass-weighted velocity dispersion for late
type galaxies is $\sigma_{\rm late}=140\,\rm{km\,s^{-1}}$ which
corresponds to a virial mass $M_\star\simeq 2.5\times10^{12} M_\sun$ within $r_v=282$\,kpc.

For simplicity we assume that the cool gas around $L^\star$ galaxies
is of the form
\begin{equation}
\rho_{\rm cool}(r) \propto r^{-2} .
\label{eq_rho}
\end{equation}
Such a density profile arises naturally in a wide range of models for
the origin of the absorbing gas. In particular, it characterizes gas
which follows the dark matter distribution in isothermal halos. It
also a property of mass conserving outflowing winds or infalling gas
(in spherical geometry) that holds over a range of radii where the
velocity is roughly constant. As we shall later see, deviations from
this power-law do not significantly affect our conclusions.

The multiple components observed in high-resolution spectra indicate
that the gas does not follow an overall smooth distribution but is
distributed in discrete systems. To reproduce this property, we
consider the cool gas to be in the form of clouds that are distributed
within the halo following a Poisson distribution with a mean number
density, $n_{\rm cloud}$ following $\rho_{\rm cool}(r)$.  Considering
a line-of-sight through the gaseous halo of a galaxy, the number of
clouds intercepted along a physical path $\d l$ is
\begin{equation}
dN_{\rm cloud}=n_{\rm cloud}({\bf r})\,\pi R_{\rm cloud}^2\, d{\bf l},
\label{dn}
\end{equation}
where $\pi R^2$ is the cloud cross-section for inducing \ion{Mg}{2}
absorption, ${\bf r}$ is a vector from the halo center, and $d{\bf l}$
is measured along the line-of-sight.  By integrating over the 
density profile of the gas we find the number of clouds along a given
line-of-sight at some impact parameter, $b$, to be
\begin{eqnarray}
N_{\rm cloud}(b)&=&\,\frac{\pi}{2} \Psi(r_v/b) 
\times n_{\rm cloud}(b)\,\pi R^2\,b
\label{ncb}
\end{eqnarray}
where $\Psi(r_v/b)=(2/\pi) \int_1^{r_v/b} d\xi
\xi^{-1}/\sqrt{\xi^2-1}$.

As mentioned above, the overall rest equivalent width of strong \ion{Mg}{2}\
systems is roughly proportional to the number of subcomponents. We can
therefore use this empirical relation and approximate the rest
equivalent width of a \ion{Mg}{2}\ absorber by
\begin{equation}
W_0 \simeq N_{\rm cloud} W_c
\label{eq_propto}
\end{equation}
with $W_c= 0.2$\AA. Here we have used the mean of the
proportionality factors reported by Bergeron \& Petitjean (1990) and
Churchill et al (2003)

The final constraint needed for our model is the normalization of the
overall amplitude.  Steidel (1993) has shown that \ion{Mg}{2}\
absorption with $W_0\gtrsim0.5$\AA\ is found within $\sim$50 kpc of
$L^\star$ galaxies with a unit covering factor.  This imposes a
constraint on our model: $\langle W_0(r=50 \rm{kpc}) \rangle \simeq
0.5$\AA . Given the previous relation, it implies that, on average,
about two \ion{Mg}{2}\ clouds are intercepted by a line-of-sight with
an impact parameter of 50 kpc.  More recent results by Zibetti et
al. (2005) \& Churchill et al. (2005) have shown that there in fact
exists a continuous distribution of impact parameters up to more than
100 kpc and the data suggest that the covering factor is less than
unity on those scales.  We note that once the function $W_0(r)$ has
been normalized at 50 kpc, the cloud-distribution of \ion{Mg}{2}\
systems introduced above automatically reproduces the drop of the
covering factor for absorption and allows absorbing clouds to be found
on scales $>$100\,kpc, as indicated by recent studies. We
note that the scatter in $W_0(b) \propto N_{\rm cloud}(b)$ is largely
due to the relatively small number of clouds. Specifically, the
standard deviation is given by $[1+\sqrt{N_{\rm cloud}(b)+0.75}]W_c$
(Gehrels 1986) and is therefore considerable ($\Delta W_0/W_0 \lesssim
1$). Interestingly, a large scatter is also seen in the data yet
its magnitude is poorly constrained due to small number statistics per
galaxy luminosity bin (Steidel et al. 1993).

We now combine all the above properties and display the resulting
cloud distribution in Fig. 1.  The left panel shows the distribution
of intercepted clouds of cool gas as a function of impact parameter
as well as the expected scatter.  The right panel presents, for
different absorption line sensitivities, the covering factor for
absorption around $L^\star$ galaxies.  We note that the observational
constraints we are using are valid on scales greater than $\sim$20
kpc.  This is denoted in Fig. 1 by the white dashed circle. On scales
smaller than 20 kpc, the contribution from the disk of late-type
galaxies becomes important. In this paper we are only interested in
gas distribution in the halo and we will focus only on these larger
scales.

\subsection{Properties of \ion{Mg}{2}\ clouds}

\begin{figure}[t]
\includegraphics[width=\hsize]{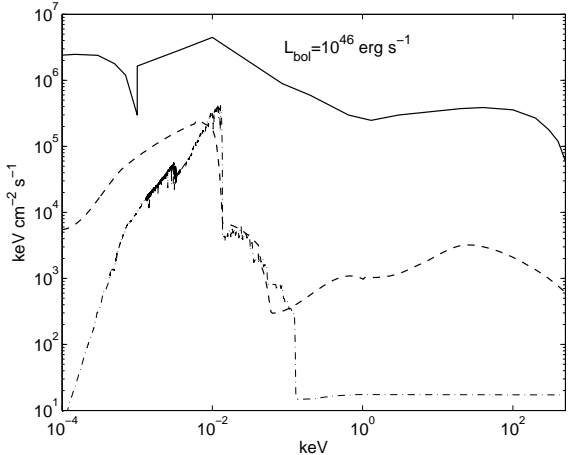}
\caption{The various components of the ionizing radiation field
considered in this work. The quasar spectral energy distribution (solid line), the galaxy
ionizing radiation field (dash-dotted line), and the UV/X-ray background
radiation field (dash line). The relative normalizations of the different
components were scaled to the values that would be measured at a distance of 50kpc  from the center of the galaxy (see
text).}
\label{sed}
\end{figure}

We now investigate the possible range of physical conditions which allow
the existence of \ion{Mg}{2}\ clouds around a galaxy.  Given the low gas
temperatures associated with individual \ion{Mg}{2}\ absorption components, we
neglect the effect of collisional heating by shocks, conduction,
and cosmic ray heating (e.g., Bergeron \& Stasinka 1986, Charlton et
al. 2000) and only consider the effects of photoionization.
For clouds in a galactic halo, most of the ionizing flux originates from the
meta-galactic field (Haardt \& Madau 1996) and the central
galaxy. Contributions to the mean ionizing photon flux from satellite
galaxies and other diffuse sources can be neglected.
The background UV/X-ray ionizing radiation is due to quasars and
stars. Characterizing this spectrum requires knowledge of the star
formation rate and quasar activation history as well as the opacity of
intervening material. In this work we use the Haardt \& Madau
(1996) calculations and consider the background radiation at $z\sim1$.
The spectrum, integrated over $4\pi$ steradians, is shown in
Fig. \ref{sed} with a dashed line.

In addition to diffuse background radiation there is also the ionizing
radiation from the central galaxy. We model the galaxy spectrum
using the Bruzual \& Charlot (1996) models and the Chabrier (2003)
initial mass function. We assume grey opacity beyond the Lyman
edge so that 5\% of the ionizing continuum escapes to the halo (e.g.,
Bland-Hawthorn \& Maloney 2001). A population of X-ray binaries is
also included (Norman et al. 2004) using the infra-red to X-ray
scaling of Ranalli et al. (2003).  The galaxy ionizing radiation is
shown in Fig. \ref{sed}.\\

To model the photoionization and thermal equilibrium of the cool gas
clouds in the halo, we use the {\sc cloudy} c06.02a photoionization
code (Ferland et al. 1998) with the Badnell di-electronic
recombination coefficients (Badnell 2006). We simplify the
calculations by assuming a slab geometry with one surface of the cloud
exposed to half of the total ionizing flux. Photoionization
calculations are carried out to the center of the cloud and symmetry
is assumed when calculating the total column density of ions (see Gnat
\& Sternberg 2004 for a more realistic treatment of the problem). We assume the
gas composition to be $0.1\times$solar (Turnshek et al. 2005, York et al. 2006).

We investigated several possible scenarios for the structure of the
clouds: (a) constant density; this is perhaps the simplest model yet
has been the main route by which a wide range of astrophysical
phenomena are modeled, (b) constant radiation and gas pressure; this
may be relevant to clouds in pressure equilibrium with their
surroundings (e.g., with the inter-cloud medium), (c) constant
temperature; this could be envisioned as gas which is collisionally
heated to the virial temperature of dark matter minihaloes. Given the
parameter space of interest, it appears that the different models do
not give appreciably different results. For clarity we will show only
the results obtained for a constant density medium.

\begin{figure}
\includegraphics[width=\hsize]{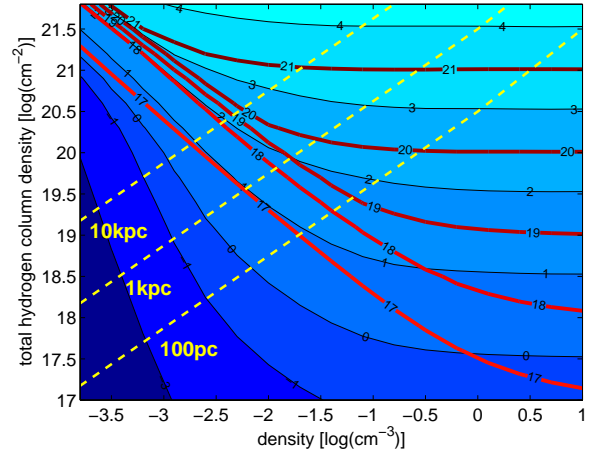}
\caption{A contour plot of the \ion{Mg}{2}\,$\lambda 2796$ optical
depth (blue contours) and the \ion{H}{1} column density (red
contours), both in logarithmic units, as a function of the cloud
particle number density and total hydrogen column density. Only the
UV/X-ray ionizing background radiation is assumed.  Lines
corresponding to several cloud sizes are also drawn (see text).}
\label{mg2_h1}
\end{figure}

We have calculated the column density of \ion{Mg}{2} and \ion{H}{1} as
a function of cloud hydrogen (ionized and neutral) column density and
the number density of particles. The results are shown in Fig.
\ref{mg2_h1}.

We have repeated the calculations including the effect of the ionizing
radiation field of the galaxy and found it to be dominant within the
central $\sim 40$\,kpc (see \S2.4 for the effect of cloud
shielding). The gas considered here, having a low ionization level, is
sensitive to the ionizing flux just above the Lyman edge.  To first
order, gas which is exposed to the galaxy ionizing flux (in addition
to the UV ionizing radiation flux, $f_{\rm UV}$), $f_{g}$ would attain
a similar ionization structure to that shown in Fig. \ref{mg2_h1}
provided its density is greater by a factor $1+f_{\rm g}/f_{\rm UV}$.
Similarly, at higher redshifts, where the UV background radiation is
stronger the density of a cloud with a similar ionization level to
that shown in Fig. \ref{mg2_h1} would be higher by the UV flux ratio
of the epochs.

\begin{figure}[t]
\includegraphics[width=\hsize]{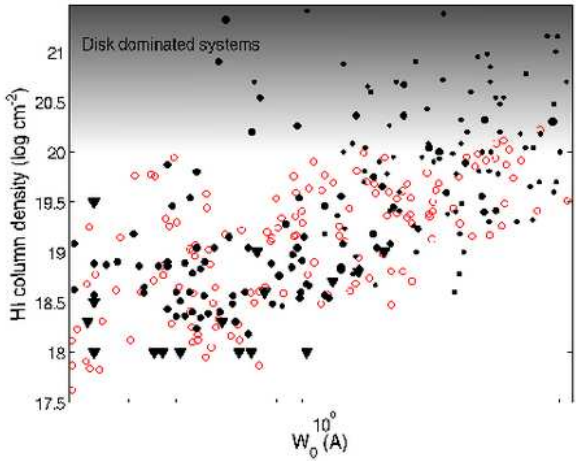}
\caption{HI column density vs. the rest equivalent width of \ion{Mg}{2}\
for \ion{Mg}{2}\ selected systems (see Rao et al. 2006).  Data appear as black
points with upper limits marked  by triangles. Different point size
indicates different selection criteria employed by Rao
et al. (large points are systems selected by criteria 1-2 while
smaller points by 3-4; see Rao et al. 2006); both sets of points have
similar properties for the low column density range considered
here. The contribution of galaxy disks to the measured H\,I column
density may become important in the shaded region of the diagram
(Zwaan et al. 2005) and our model does not attempt to account for it.
The simulated data are drawn in red points and reproduce the lower
column density envelope (see text).}
\label{rao}
\end{figure}

Constraining the parameter range relevant to strong \ion{Mg}{2}\
absorbers is not straightforward since strong \ion{Mg}{2}\ lines are
usually saturated and additional observational constraints are needed
to estimate their typical density and column density. A useful
constraint is the size of the clouds derived from strong lensing
studies. As noted above, such studies indicate that the size of
\ion{Mg}{2}\ absorbers is $\sim$1\,kpc. This constraint substantially
reduces the relevant parameter space in Fig.  \ref{mg2_h1}. A second
constraint comes from the \ion{H}{1}\ column density measurements for
strong absorbers (Rao et al. 2006). Using the rest equivalent width as
a proxy for the column density along the line-of-sight (see
eq. \ref{eq_propto}) we see from Fig. \ref{rao} that a typical cloud
has an \ion{H}{1}\ column density of $\sim 10^{18}~{\rm cm^{-2}}$,
implying a density of about $2\times 10^{-2}~{\rm cm^{-3}}$. Such
values are consistent with previous estimates in the literature for the density of cool gas in galaxy halos (e.g.,
Sternberg, McKee, \& Wolfire 2002 for the case of high velocity clouds in the Galaxy).

The scatter in the $\mathrm{N_{HI}}$ vs. $W_0$  relation presented in Fig.
\ref{rao} is significantly larger than that of the simplified model
presented in the previous section.  Several effects not included in
our model could also contribute to this scatter: a distribution of
galaxy masses, types, star formation histories, etc. -- all of which
are beyond the scope of this paper and are not observationally
motivated at this stage. In the absence of a clear physical
motivation, we attempt to take this effect into account by introducing
a distribution of cloud sizes (as implied by the observations; \S
2.1). Here, clouds have similar densities which allow them to be in
pressure equilibrium with their environment. (We find the alternative
scenario in which all clouds have the same size but different
densities less plausible since the implied evaporation times would be
embarrassingly short compared to the Hubble time; see appendix.) To
this end we assume that the size of individual Mg\,II clouds is drawn
from an intrinsic distribution $\phi(R_{\rm cloud})$ so that observed
one (taking into account cloud cross-section effects on the
detectability rates) $\phi_{\rm obs}(R_{\rm cloud})\propto\phi(R_{\rm
cloud})\times R_{\rm cloud}^2$. A simple power-law distribution gives
a reasonably good description of the data:
\begin{eqnarray}
\phi(R_{\rm cloud}) \propto R_{\rm cloud}^\beta\,,  {\rm ~~with~~} R_{\rm
cloud}^{\rm min}<R_{\rm cloud}<R_{\rm cloud}^{\rm max} \,,
\end{eqnarray}
which implies a mass distribution:
\begin{eqnarray}
\phi(M_{\rm cloud}) \propto M_{\rm cloud}^{(\beta-2)/3}\,, {\rm
~~with~~} M_{\rm cloud}^{\rm min}<M_{\rm cloud}<M_{\rm cloud}^{\rm max}
\end{eqnarray}
where $M_{\rm cloud}$ is the mass of a single cloud. This distribution
is defined between some minimum and maximum sizes. We generate a
population of clouds as follows: in each rest-equivalent width bin we
randomly picked the sizes of $N_{\rm cloud}$ clouds drawn from $\phi$
and calculated their ionization structure and their \ion{H}{1}\ column
densities. The total \ion{H}{1}\ column of all clouds was then summed
taking into account the scatter caused by intercepting a spherical
cloud at various positions. This procedure resulted in one simulated
point in Fig. \ref{mg2_h1} which was then repeated covering the entire
rest-equivalent width range spanned by the data.  As we are only
interested in modeling the gas contribution originating from the halo
of a galaxy, we do not model the high HI column density values with
$N_{\rm HI}>10^{20} \rm{cm}^{-2}$ and for which a disk is a likely
contributor (Zwaan et al. 2005). We note that our estimates for the
total mass of cool gas in the halos of galaxies will change by a
factor of order unity by including the upper end of the column density
distribution in our model. Given our density estimates from
photoionization calculations and the constraints on the mean size of
clouds then a good fit for the Rao et al. dataset is obtained for
$R_{\rm cloud}^{\rm min}=0.3$\,kpc, $R_{\rm cloud}^{\rm max}=1.5$\,kpc
and $\beta=-2.5$ (corresponding to $\beta=-0.5$ for the observed
distribution). Our approach allows us to reproduce the main
observational trends summarized in Fig. \ref{rao}.

The gas temperature in our model is $\sim 10^4$\,K and so the gas
pressure $P/k_B\sim 200~{\rm cm^{-3}\,K}$ (i.e., similar to the value
used by e.g., Mo \& Miralda-Escude 1996 and Gnat \& Sternberg 2004 but
somewhat lower than that implied by Fukugita \& Peebles 2006).  The
implied cloud mass distribution, $\phi(M_{\rm cloud})\propto M_{\rm
cloud}^{-1.5}$, with $M_{\rm cloud}^{\rm min}=6\times 10^4~{\rm
M_\sun}$ and $M_{\rm cloud}^{\rm max}=7\times 10^6~{\rm M_\sun}$.
Clouds are Jeans stable provided that the total mass (dark and
baryonic matter) is $M_{\rm cloud}<10^8~{\rm M_\sun}$.  The clouds are
also relatively stable with respect to hydrodynamic instabilities
(such as Klevin-Helmholtz) and evaporation (Maller \& Bullock 2004). 

It is interesting to note that our typical \ion{Mg}{2} absorbing
cloud hydrogen is mostly ionized.  Most of the metals are in singly or
doubly ionized configuration. Nevertheless, the large total gas columns
result in higher ionization levels having a non-negligible column
density. Specifically, the column density of
\ion{C}{4} in our model can exceed $10^{13}~{\rm cm^{-2}}$ which
translates to an optical depth of order unity and is therefore
consistent with a picture in which \ion{C}{4}\ and \ion{Mg}{2}\
absorption have a common origin. The uniform density, cool clouds 
considered here cannot account for the strong intervening \ion{O}{6}\,$\lambda 
1035$  systems (see Danforth \& Shull 2005, Bergeron \& Herbert-Fort 2005) and
an additional component/physics  is required to explain them (e.g., Mo \& Miralda-Escude  
1996, McKee \& Begelman 1990 and references therein).

In our analysis we have assumed that all clouds are exposed to a
similar flux level regardless of their location in the
halo. Nevertheless, the effect of shielding of the ionizing flux by
other clouds could be, in principal, important. To estimate the magnitude
of this effect let us consider a cloud located 50\,kpc from the
galaxy. Such a cloud would see an effective column density of a
few\,$\times 10^{18}~{\rm cm^{-2}}$ (corresponding to a few
clouds). When shielding is included as well as the contribution from
the central galaxy, we find an overall decrease in the photoionization
rate by a factor $\sim 2$ compared to the non-shielded case.

\subsection{Summary \& limitations of the model}

As stated above, our model is aimed at providing a
framework for studying the effects of a quasar radiation field on the
absorption and emission properties of the corresponding system.  We
have calibrated it using observational constraints on the spatial
distribution of \ion{Mg}{2}\ clouds within the halo of $\sim L^\star$ galaxies
and the corresponding HI column densities, and using photoionization
calculations we have shown that the implied cloud parameters are
consistent with the existence of \ion{Mg}{2}\ gas. We summarize the main model
parameters in Table \ref{table_summary}.

\begin{figure}
\begin{center}
\includegraphics[width=\hsize]{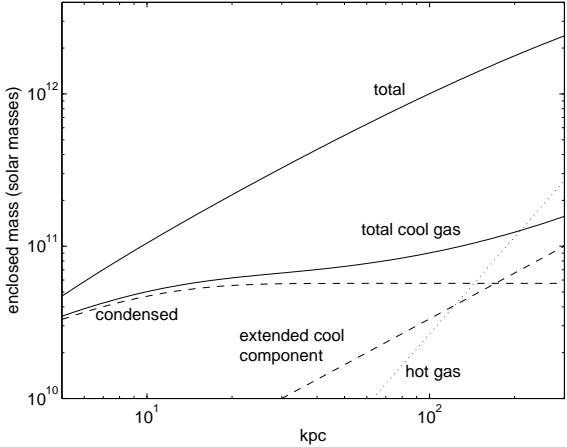}
\caption{The extended cool gas component considered in this work
within the framework of the massive corona model by Fukugita \&
Peebles (2006; see their paper for the assumptions and physical
motivation behind each component). The total mass in cool gas is
comparable to that of the hot component within the virial
radius. Given the condensed form of the extended cool component it
occupies only a small fraction ($\sim 1\%$) of the entire volume.}
\end{center}
\label{fig_masataka}
\end{figure}

\begin{table}[!t]
\begin{center}
\caption{Model summary}
\begin{tabular}{ll}
  \hline\hline
~~~~~~~~~~~~~~Cloud parameters&\\
\hline
Gas mass & $6\times10^4<M_{\rm cloud}<7\times10^6 M_\sun$\\
Radius & $0.3<R_{\rm cloud}<1.5$ kpc\\
Density & $\rho_{\rm cloud}=0.02$ atom cm$^{-3}$\\
Metallicity & 0.1 $Z_\sun$\\
\hline\hline
~~~~~~Halo parameters ($20\, \rm{kpc}<r< r_v$)&\\
\hline 
Cool gas density profile & $\rho\propto r^{-2}$\\
Dark matter velocity dispersion & $\sigma=140$ km/s\\
Virial radius & $r_v=282$ kpc\\
Virial mass & $M\simeq10^{12} M_\sun$ \\
Cool gas Mass\footnote{for a covering factor of unity and within $r_v$ (see section 2.4)}  & $M_{\rm cool}\sim10^{11} M_\sun$ \\
Number of clouds\footnotemark[a] & $\sim7\times 10^4$ \\
\hline\hline
\end{tabular}
\label{table_summary}
\end{center}
\end{table}

This model leads to a picture in which $L^\star$ halos are filled
with gaseous clouds which are partially ionized and sufficiently
optically thick to be detected in both low- and high-ionization (e.g.,
\ion{C}{4}) absorption lines. These clouds have sizes of order 1\,kpc,
masses of $\sim10^6 {\rm M}_\sun$ and particle densities $\sim
10^{-2}\,\rm{cm^{-3}}$, amounting to a cool gas mass of $\sim 10^{11}
{\rm M}_\sun$ \emph{within the virial radius}, i.e. $\sim 280$ kpc.

The results presented above assume that cool gas clouds are
distributed throughout the entire volume of the halo up to the virial
radius. Nevertheless, the virial radius has no clear physical
connection to the scale, $r_{\rm cool}$, up to which cool clouds
exist. In particular, the value of $r_{\rm cool}$ is expected to
depend on the (unknown) origin of the cool gas. For example, if the
gas is ejected from the galaxy $r_{\rm cool}$ is likely to be smaller
than the case in which cool gas is accreted from the inter-galactic
medium. Moreoever, different values may be obtained if the cool gas
condenses out of the virialized halo gas (e.g., Mo \& Miralda-Escude
1996). At present, little is known with confidence about the origin of
such cool gas and observations provide only very loose constraints,
hence $r_{\rm cool}$ is ill-determined.  That being said, we find that the
mass of cool gas in galaxy halos depend relatively little on $r_{\rm
cool}$ for the relevant scale range. Specifically, to reproduce the
covering factor statistics on large scales, we obtain $M_{\rm
cool}\sim 3\,(4)\times 10^{10}\,{\rm M_\odot}$ for $r_{\rm
cool}=50\,(100)$\,kpc. Clearly, the mass of cool gas in galaxy halos
is considerable for all plausible values of $r_{\rm cool}$ and a
covering factor of order unity.  In what follows we therefore assume
the gas occupies the entire virial volume (i.e., $r_{\rm cool}=r_v$).

We note that, given our model assumptions and calibration constraints,
the overall mass of cool gas, $M$, scales with the mean cloud size,
$\left < R_{\rm cloud} \right >$ and is inversely proportional to the
assumed gas density.  It must be emphasized that the total mass of
cool gas depends on the assumed covering for absorption, i.e. a
quantity still debated in the literature.  In the present study we
have used a covering factor of unity for strong MgII absorption within
$\sim 50$ kpc, as suggested by Steidel et al. (1997). However, other
studies claim lower values ($\sim0.2$ from \citet{1992ApJ...396...20B}
and $\sim 0.5$ from {Tripp
\& Bowen 2005}). As the total mass of cool gas scales linearly with
this quantity, we might be overestimating the mass if the actual
covering factor is not unity.

According to our model, the mass of cool gas in galaxy halos scales
linearly with halo size.  Specifically, the mass within 30\,kpc is
$M_{\rm cool} < 10^{10}\,{\rm M_\odot}$ and, by extrapolation to
smaller scales, is comparable to that estimated by Martin (2006) for
ultra-luminous infrared galaxy outflows.  In Fig. \ref{fig_masataka}
we show the different components of the average mass profile of
late-type galaxies proposed by Fukugita \& Peebles (2006), and show
that including the distribution of cool gas proposed in the present
paper is only a modest departure from it.

\begin{figure}
\includegraphics[width=\hsize]{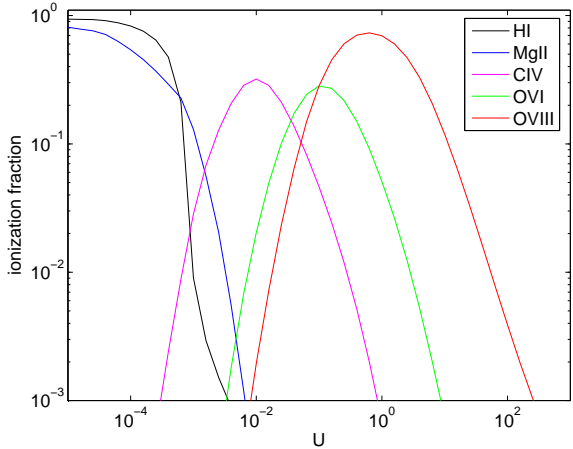}
\caption{The ionization fraction of the abundant elements as a
function of the ionization parameter, $U$. In this case the radiation
field of the quasar overwhelms all other sources within $\sim 10Mpc$
of the quasar. Gas with the same density and distance as a
typical \ion{Mg}{2}\ absorbing cloud would be highly ionized.}
\label{quasar}
\end{figure}

\begin{figure*}[ht]
\plottwo{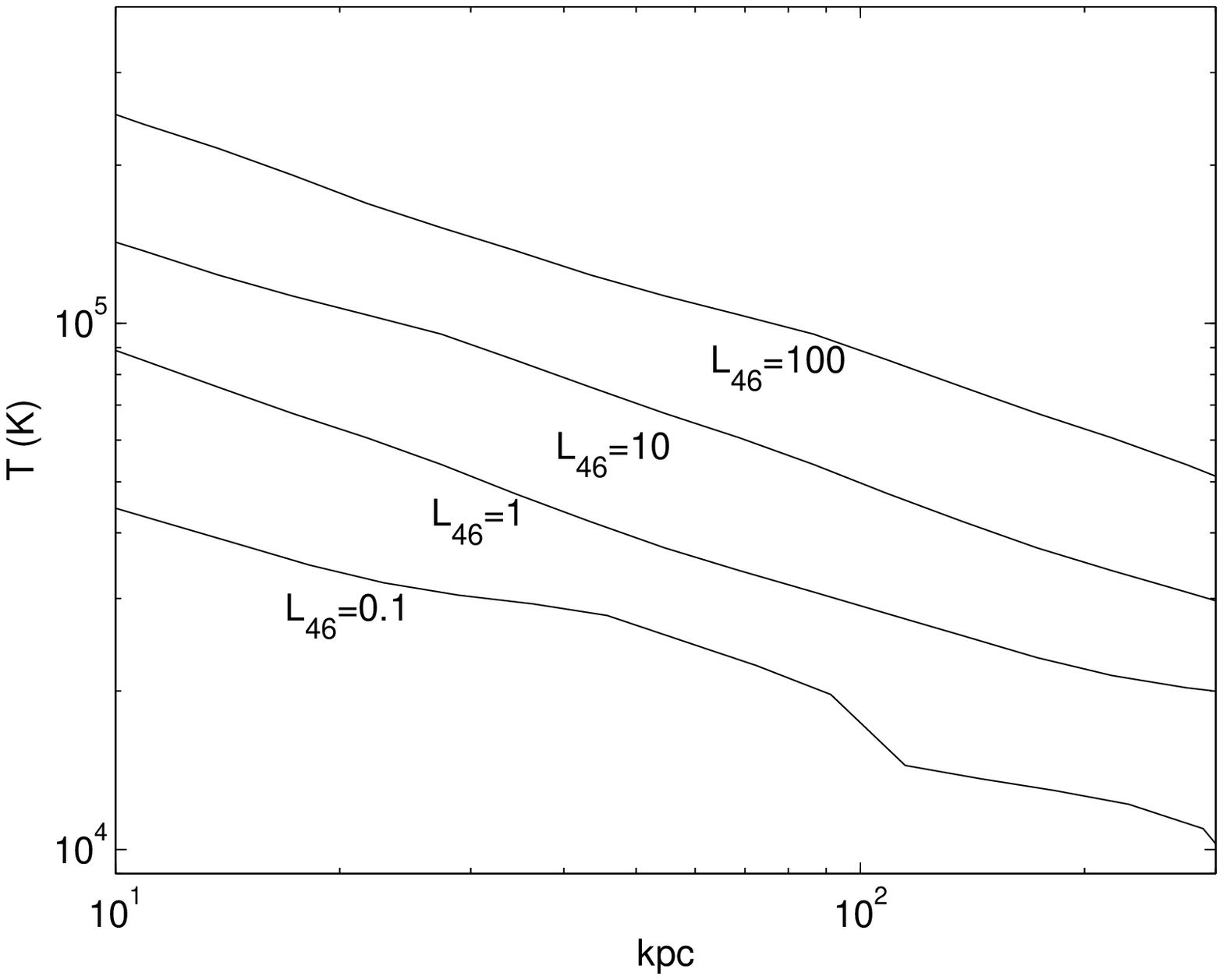}{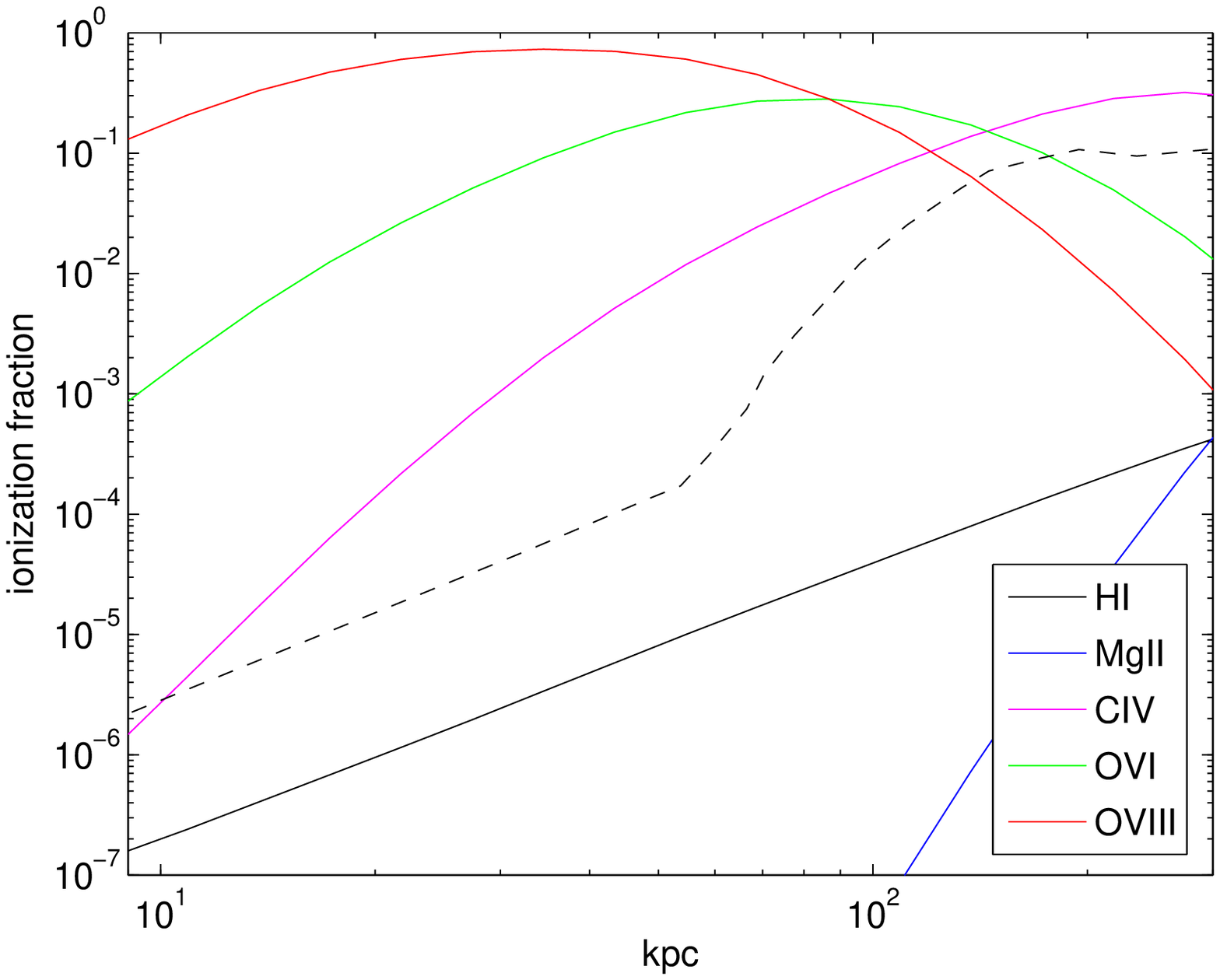}
\caption{The thermal and ionization properties of halo gas ($f_M=5$) illuminated by a quasar. {\it Left:} The temperature profile for a constant density, optically
thin medium with density of $0.02\,{\rm cm^{-3}}$ (see text) for
several values of quasar luminosity. Clearly, the temperature drops
with distance as the ionization parameter decreases ($\propto
r^{-2}$). The radiative effect of Seyfert galaxies on their
environment is more localized than that of bright quasars.  {\it
Right:} The ionization structure as a function of distance from a
$L_{46}=1$ quasar. As expected, the ionization level decreases with
distance due to the dilution of the radiation field. The fraction of
\ion{Mg}{2}\ recovers only on Mpc scales while closer to the quasar
(on $\sim 100$\,kpc scales) \ion{C}{4}\ is expected to show absorption
signatures while on smaller scales still ($\sim 20$\,kpc) \ion{O}{6}\ would
show up in the spectrum. The gas is very highly ionized within $\sim
30$\,kpc and could be detected in the X-rays. For quasars with
different luminosities the abscissa should be rescaled by
$L_{46}^{1/2}$. Thus, brighter quasars would ionize their environment
to larger distances. The dashed line corresponds to the ionization
fraction of \ion{H}{1}\ for the case of $L_{46}=0.1$. The hump beyond $\sim
50$\,kpc is the result of self-shielding (see text).}
\label{TIvsD}
\end{figure*}

So far we have concentrated on modeling the gas within the halo of an
$L^\star$ galaxy with $M_{vir}=2.5\times 10^{12}\, {\rm M}_\sun$.  It
is interesting to note that our model \emph{may} be scaled to
different galaxy luminosities and masses.  Indeed, analyses by Steidel
et al. (1997) and Guillemin \& Bergeron (1997) have indicated that the
normalization of the \ion{Mg}{2}\ rest equivalent width follows a
relation
\begin{equation}
W_0(r,L)=W_0(r)\times
\left( \frac{L}{L^\star} \right)^\gamma
\label{sca}
\end{equation}
with $\gamma\simeq 0.1-0.4$.  Such a scaling
has been reported in the luminosity range $0.3-5 L^{\star}$ and at
redshifts $0.5<z<1$. Extrapolating it further is therefore
uncertain. The above relation may be understood within the framework
of our model as a dependence of the number of clouds, hence mass of
cool gas, on luminosity. Denoting the mass of cool gas in $L^\star$
galaxies by $M^\star$ then the mass of cool gas in halos of galaxies
with luminosity $L$ may be written as $M(L)=f_M(L) M^\star$ where $f_M$ is a dimensionless scaling factor.

\section{Quasar haloes}
\label{section_quasar}

We now investigate the effects of a quasar on the gas distribution
within a halo and discuss the observational signature in absorption
and emission.  Our main assumption is that quasar halos are analogous
to those of non-active galaxies. As indicated by recent studies (e.g.,
Jahnke et al. 2004), the hosts of bright quasars seem to correspond to
a few$\times L^\star$ galaxies. Furthermore, a recent study by Serber
et al. (2006) seems to indicate that bright quasars reside in
over-dense regions implying that their dark matter halos are several
times those typical of $L^\star$ galaxies. Motivated by these studies
we shall allow $f_M$ to deviate from unity. We  show below that
many observational constraints may be accommodated by using $f_M\sim
5-10$ for the mass of cool gas in quasar halos.

\subsection{Thermal and Ionization Structure}


In what follows we compute the time-{\it independent} ionization and
thermal structure of a cloud exposed to a constant quasar ionizing
flux. In the Appendix we present time-{\it dependent} calculations and show
that the steady-state assumption suffices for the luminosity range and
the relevant spectral features considered here.

In modeling the spectral energy distribution (SED) of the quasar we
have used the continuum defined by Sazonov, Ostriker, \& Sunyaev
(2004) which is shown in Fig. \ref{sed}.  Ionizing sources other than
the quasar can be safely neglected. We start by introducing
the ionization parameter, $U$, which is the ratio of photon density
to particle number density and is a measure of the ionization
level for photoionized gas. For the chosen SED we obtain that
\begin{equation}
U\simeq 0.6\, L_{46}\,\left ( \frac{n}{10^{-2}~{\rm cm^{-3}}} \right
)^{-1} \left ( \frac{r}{50~{\rm kpc}} \right )^{-2}.
\label{ion_param}
\end{equation}
where the quasar luminosity is $L=10^{46}L_{46}~{\rm erg~s^{-1}}$.
The ionization structure of several commonly detected ions as a
function of $U$ is shown in Fig. \ref{quasar}. By combining these
results with the density estimates from our model (\S 2), we can compute the ionization level and temperature
as a function of the radius from the center of the galaxy.
The results for the temperature profile are shown in Fig.
\ref{TIvsD}. As expected, the quasar has a substantial effect on the
thermal state of the inner parts of the halo and can heat the
(initially cool) gas to high temperatures by means of
photo-absorption.  The extent to which the \ion{Mg}{2}\ halo is
ionized depends mainly on the quasar luminosity. For example, a quasar with
$L_{46}=1$ heats gas within the inner 20\,kpc to some $10^5$\,K with
gas at 50\,kpc being ionized to a few\,$\times 10^4$\,K. Objects with
Seyfert-like luminosities ($L_{46}\lesssim0.1$) have a smaller effect
on their immediate environments and heat the gas within their inner
20\,kpc to a few\,$\times 10^4$\,K with clouds at larger radii
remaining relatively cool. In contrast, very luminous quasars can heat
their 100\,kpc environment to temperatures of order $10^5$\,K with gas
within the inner 10\,kpc being able to reach the Compton temperature.
The effect of a luminous quasar is not necessarily limited to halo gas
and low density gas may be affected on much larger (Mpc)
scales. Our calculations indicate (Fig. \ref{quasar}) that for
clouds to remain relatively cool within the central 30\,kpc of a
$L_{46}=1$ source, their density should be $\gg 10^2~{\rm cm^{-3}}$,
i.e. considerably higher than the values considered here and those thought
to be representative of
\ion{Mg}{2}\ clouds (c.f., Stockton et al. 2002).

\begin{figure}
\includegraphics[width=\hsize]{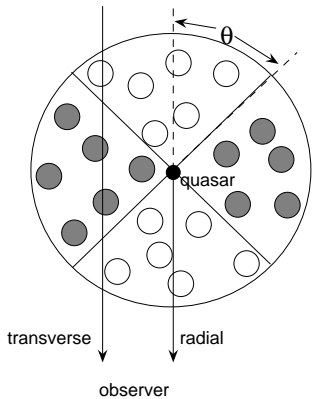}
\caption{The geometry considered here whereby a quasar emits ionizing radiation into a cone with opening angle $\theta$. Cool gas clouds which are exposed to the ionizing radiation (circles) are heated while those that are shielded from the ionizing radiation remain cool (filled circles) and their properties resemble those of halos associated with non-active galaxies. Off-center lines of sight will intercept cool gas in the transverse direction, while those that point radially toward the quasar will mostly see ionized gas (see text).}
\label{geometry}
\end{figure}

\begin{figure}
\plotone{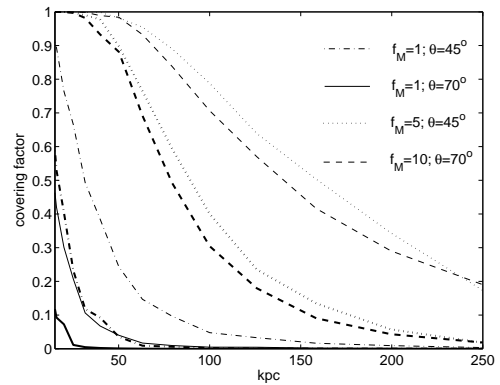}
\caption{The covering factor for strong \ion{Mg}{2}\ absorption as a function of the impact parameter for lines-of-sight passing through the halo of a quasar (see figure \ref{geometry}). Thick (thin) lines correspond to the covering factor for systems with $W_0>1$\AA\ ($W_0>0.6$\AA). Four halo models are considered that differ by the mass of cool gas within the virial radius (denoted by $f_M$; see text) and the opening angle, $\theta$, of the quasar radiation field. As expected, the covering factor declines with increasing impact parameter due to the decreasing cloud density with distance. The covering factor for strong absorption is sensitive to both $\theta$ and $f_M$. As $\theta$ increases, more gas is ionized and the halo becomes transparent to \ion{Mg}{2}\ absorption. As $f_M$ rises, more clouds are intercepted along lines-of-sight through the halo. As shown, different combinations of $f_M$ and $\theta$ can result in a similar covering factor for strong \ion{Mg}{2}\ absorption (see text). }
\label{mg2_transverse}
\end{figure}

The ionization structure closely follows the temperature
profile. Fig. \ref{TIvsD} shows the ionization fraction of several
ions as a function of radius in the halo for the case of
$L_{46}=1$. Clearly, gas within the inner 100\,kpc is devoid of
\ion{Mg}{2}, with \ion{C}{4}\, \ion{O}{6}, and \ion{O}{8}\ being most
abundant; the ionization fraction of the latter peaking on the
smallest scales.  For a $f_M=5$ halo with a $L_{46}=1$ quasar (e.g.,
Jahnke et al. 2004 and equation \ref{sca}), the gas remains optically
thin above the Lyman edge. For different source luminosities the
abscissa should be rescaled by $L_{46}^{-1}$. For low quasar
luminosities ($L_{46} <0.1$) even the inner regions of the halo remain
at low temperatures and the gas is less affected by the quasar on
$\sim 100$\,kpc scales (see Fig.
\ref{TIvsD}).\\

 The active galactic nuclei (AGN) unification scheme (Antonucci 1993)
 suggests that quasars emit light in a double cone geometry with an
 opening angle $\theta$ (see Fig. \ref{geometry}).  Such an
 anisotropy in the radiation field will be imprinted in the properties
 of the gas distribution around the quasar. In the next sections we
 discuss the observational implications, both in absorption and
 emission.

\subsection{Quasar-halo absorption properties}

Clouds of gas that lie within the ionization cone of the quasar would be heated
and ionized, and would follow the ionization structure presented in
Fig. \ref{TIvsD}. Therefore, the line-of-sight to a quasar would be
largely devoid of associated Mg\,II absorption due to the halo clouds
modelled in this paper. Nevertheless, such highly ionized clouds are expected
to give rise to associated absorption features from species like H\,I,
C\,IV, and O\,VI -- depending on their ionization structure which is
determined by the quasar luminosity and their distance from the source. In contrast, lines-of-sight
probing the quasar halo at finite impact parameters from the quasar
would intercept cool gas which is not affected by the quasar radiation
field and is therefore likely to show absorption features due to low
ionization species as those seen around non-active galaxies. The
number of \ion{Mg}{2}\ clouds intercepted by a given sight-line depends not
only on the total number of clouds within the halo (which, for
non-active galaxies, depends on the galaxy luminosity; see equation 9)
but also on the opening angle of the quasar.

Given the above model, we now discuss the observable properties of
quasar haloes in terms of their expected absorption signatures.  In a
recent work we have used pairs of quasars at different
redshifts and investigated the presence of \ion{Mg}{2}\ absorption around the
foreground objects (Bowen et al. 2006). In 4/4 cases we have detected
\ion{Mg}{2}\ absorption with $W_0>0.5$\AA\ in the spectrum of the background
quasars with a redshift matching that of the foreground quasar.  This
suggests that strong \ion{Mg}{2}\ absorbers are common in the
transverse direction of quasars.  We have extended our analysis to a
larger sample of systems and our new results (Bowen et al. 2007)
indicate that the presence of strong \ion{Mg}{2}\ absorption around
quasars with a covering factor close to unity within 100 kpc, with
a rapid decline on larger scales.  Such behavior is similar to that
of \ion{Mg}{2}\ gas which is detected around galaxies. Interestingly,
the high covering factor for \ion{Mg}{2}\ gas in the transverse
direction seem to be in contrast with smaller values ($\sim 10\%$;
Aldcroft et al. 1994) derived for \ion{Mg}{2}\ absorption along the
line-of-sight to these objects (see Fig. \ref{geometry}). As we
show below, this surprising finding is naturally explained by our
model.

Quasar radiation is thought to be emitted into cones of opening angle
$\theta$ (Antonucci 1993). As such, $\theta$ is a measure of the mass
of cool gas within the halo which is exposed to the ionizing field and
determines the mass of the remaining cool gas. The connection between
$\theta$ and $W_0$ is therefore apparent within the framework of the
model (see figure \ref{geometry}).  Model predictions for the
absorption covering factor as a function of the rest equivalent width,
the impact parameter, and the mass of cool gas, $f_M$, are given in Fig.
\ref{mg2_transverse}. These results are analogous to those of galaxies
but for the additional dependence on $\theta$. As shown, quasars that emit more isotropically have less
cool gas at any impact parameter due to ionization effects. More
massive envelopes of cool gas have a larger number of clouds and so result in broader \ion{Mg}{2}\ troughs. Clearly,
different combinations of $f_M$ and $\theta$ can yield similar covering
fractions. At present, neither the mass of cool gas
in quasar halos nor the opening angle of quasar ionization cones are
well determined with recent surveys indicating that of $\theta$ may
depend on the quasar luminosity and is in the range $45^\circ
<\theta<70^\circ$ for bright quasars (Willott et al. 2000, Treister \&
Urry 2005). Fig. \ref{mg2_transverse} shows that $f_M\simeq 5-10$ is
required to explain the high occurrence of transverse strong
\ion{Mg}{2}\ absorption around quasars within 100 kpc in the Bowen et
al. (2006, 2007) sample.

\begin{figure}
\plotone{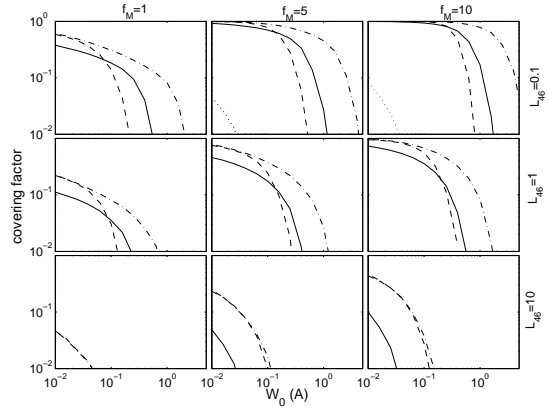}
\caption{The predicted covering factor for associated absorption where lines-of-sight probe the ionized part of the halo. Covering factors for several high ionization lines are considered correspnding to: \ion{H}{1}\,$\lambda 1216$ (dash-dotted line), \ion{Mg}{2}\,$\lambda 2800$ (dotted line), \ion{C}{4}\,$\lambda 1548$ (solid line), and \ion{O}{6}\,$\lambda 1035$ (dashed line). The results are shown as a function of the quasar luminosity for $L_{46}=0.1-10$ (columns) and mass in cold gas with $f_M=1-10$. As shown, the covering factor is more sensitive to the luminosity rather than the mass in cold gas.  The covering factor follows the exponential cutoff of the
Poisson distribution for the respective lines. Low luminosity quasars
have a higher covering factor for low ionization lines since the halo
is less ionized. For bright objects the halo is more ionized and
low ionization lines become weaker.}
\label{associated_abs}
\end{figure}

We can further compare the predictions of our model to observations by
considering the fraction of quasars showing associated absorption,
i.e. absorption arising along the quasar line-of-sight close to the
quasar redshift (Fig. \ref{geometry}).  As mentioned above, in such a
case, low-ionization species are not expected to be abundant due to
rapid ionizations by the quasar radiation field. However, the same
clouds are expected to give rise to higher ionization absorption
lines. In Fig. \ref{associated_abs} we present predictions for the
covering factor of \ion{H}{1}, \ion{Mg}{2}, \ion{C}{4}, and \ion{O}{6}
lines as a function of halo mass and quasar quasar luminosity. The
covering factor for all lines is seen to decrease exponentially with
increasing $W_0$ due to Poisson statistics. As shown, the covering
factor for absorption is more sensitive to the quasar luminosity than
to the mass of cool gas in the halo. In particular, the halos of more
luminous quasars are more ionized resulting in a lower covering factor
for strong low ionization systems. As expected, more massive halos
produce stronger absorption due to the larger number of clouds
intercepted along the line-of-sight. Predictions appear to be in
qualitative agreement with recent surveys indicating a covering factor
of $\sim 30\%$ for strong \ion{C}{4}\,$\lambda \lambda 1548,1550$
absorption (Vestergaard et al. 2003). Our model predicts little
associated
\ion{Mg}{2} absorption due to halo gas.  We note however, that our
model does not include the, possibly important, contribution from
absorption by gas in the immediate vicinity of the black hole
(sometimes termed intrinsic absorption). That said, our model is
consistent with the notion that most low ionization associated systems
are intrinsic to the quasar. Additional data concerning the covering
factor of various ionization stages as a function of quasar luminosity
(and perhaps host mass) are required to test our model.

We note that a complementary configuration to the one used in the
Bowen et al. (2006, 2007) survey is provided by radio galaxies. For
these objects our line-of-sight passes outside the quasar radiation
cone. Measurements of large-scale \ion{H}{1}\ absorption by Van Ojik
et al. (1997) have shown that typical column densities through the
neutral part of the halo are of order $\lesssim 10^{19}~{\rm
cm^{-2}}$. Such values are in qualitative agreement with our model and
with the Rao et al. (2006) dataset (see \S 2). A more detailed
comparison is unwarranted at this stage.

\subsection{Quasar-halo emission properties}

As the gas is photo-excited by the radiation field of the quasar,
clouds would not only absorb but also emit and scatter continuum and
line photons. In this section we calculate the expected emission line flux from
the halo.

\begin{figure*}
\includegraphics[width=.49\hsize]{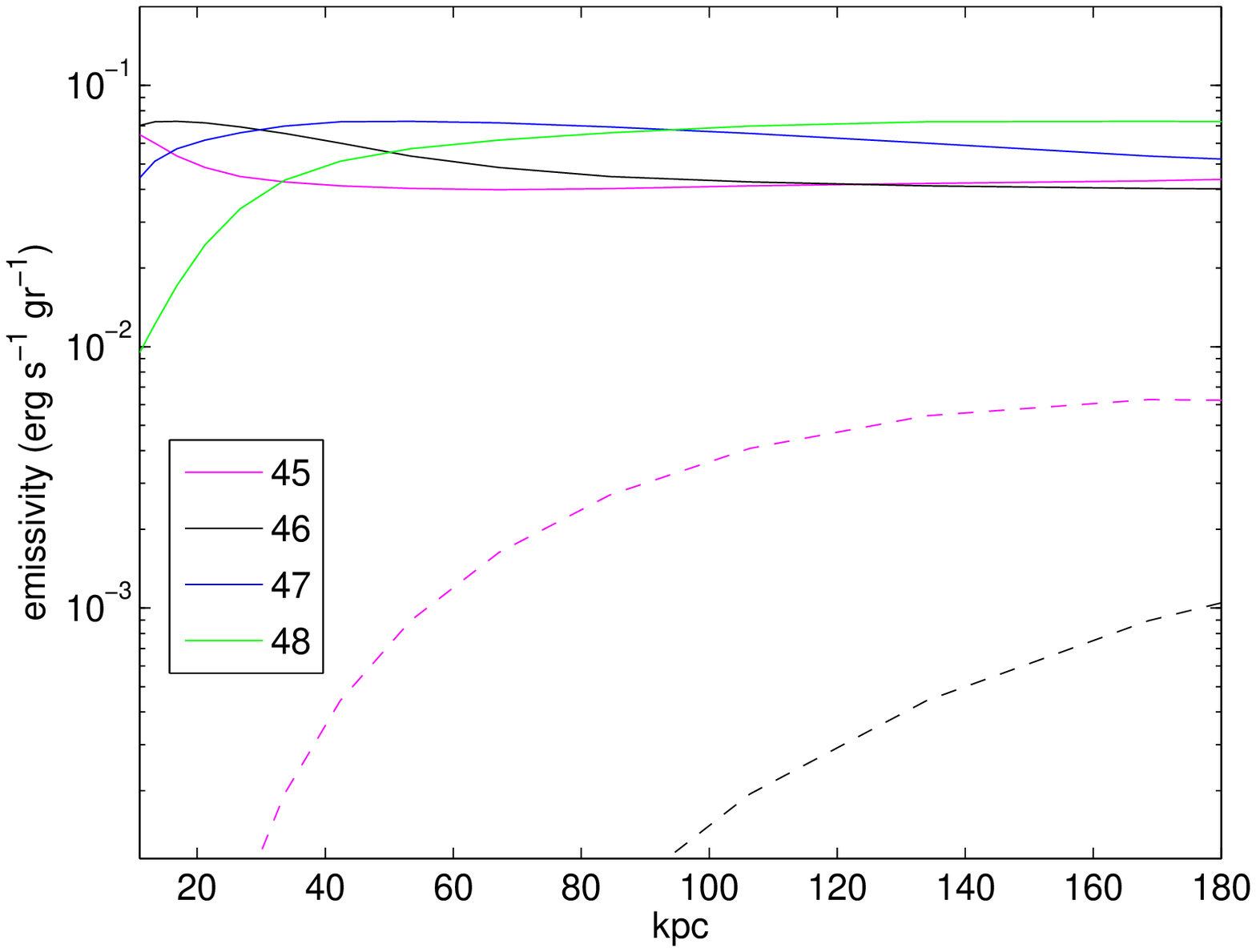}
\includegraphics[width=.49\hsize]{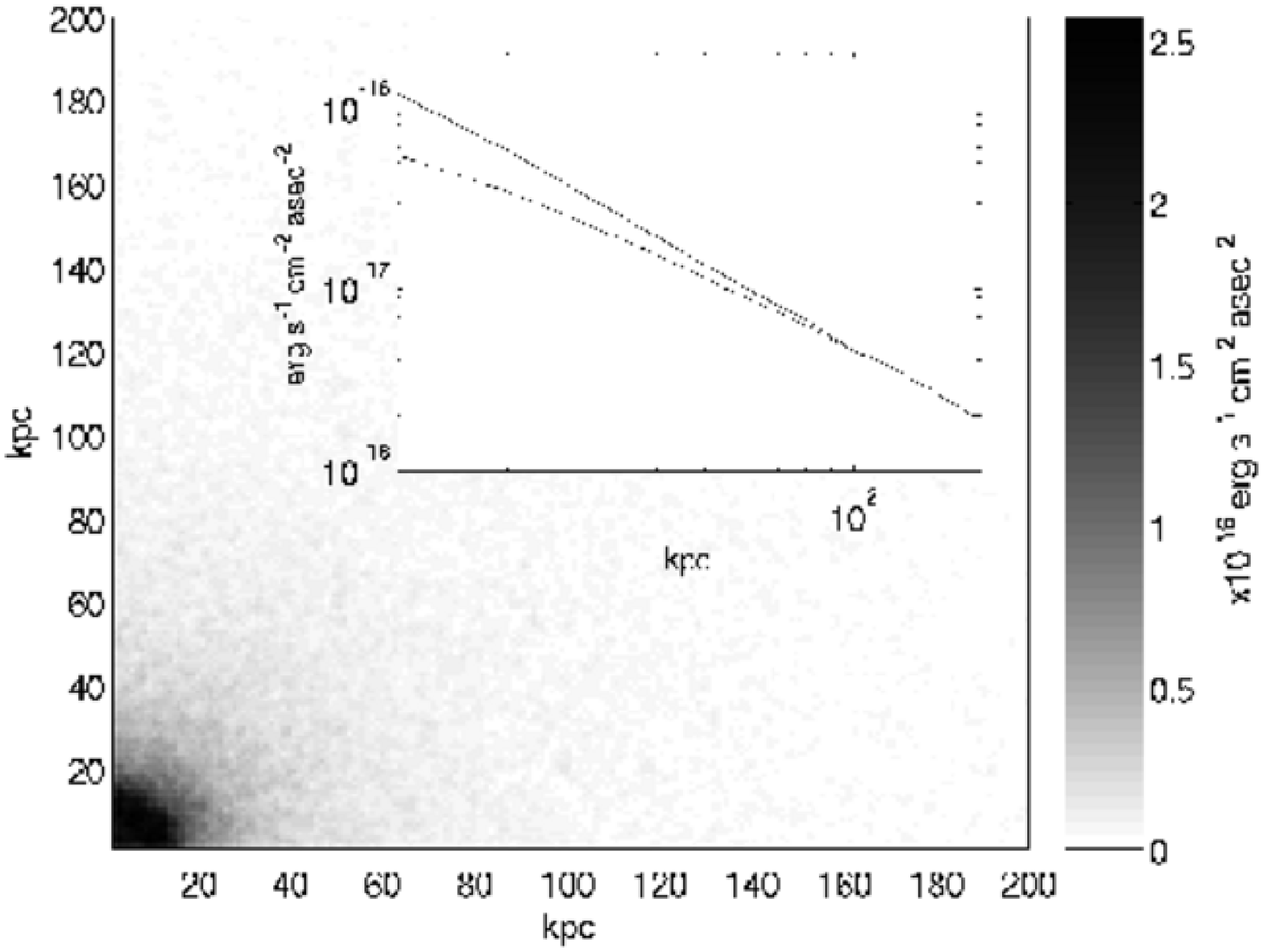}
\caption{The Ly$\alpha$ (solid) and [\ion{O}{3}]\,$\lambda 5007$ (dashed)
emissivities for various color coded luminosities. While the Ly$\alpha$
emissivities of the nebulae at large radii are similar, the emissivity
decreases with increasing quasar luminosity on smaller, $\lesssim
30$\,kpc scales due to ionization effects. The [\ion{O}{3}]\,$\lambda 5007$ emissivities
are lower than those of $L\alpha$ and have their maxima at larger
radii where cool gas is more abundant.
\emph{Right:} The Ly$\alpha$ halo surrounding an $L_{46}=1$
($f_M=5,~\theta=70^\circ$) quasar at $z=1$ with colors reflecting the flux
level. Also shown in an inset is the surface brightness profile
including (solid line) and excluding (dashed line) scattering of
Ly$\alpha$ photons from the broad emission line region of the quasar
(see text).}
\label{lumi_lumi}
\end{figure*}

Line emissivity depends on the ionization state of the gas as well as
on the optical depth through which line photons need to propagate,
escape, and reach the observer. In calculating the emissivity of lines
we have assumed that once line photons escape individual clouds, they
reach the observer; i.e., we assume no photon scattering or absorption
among different clouds. This approximation is adequate if clouds along
our line-of-sight have little overlap in velocity space, which is the
case considered in this paper. The fraction of line photons escaping
from a single cloud was calculated using {\sc cloudy}  using the escape
probability method for a static cloud. The resulting emissivity per
unit mass as a function of distance from the ionizing source is shown
in Fig. \ref{lumi_lumi} for Ly$\alpha$ emission line. As shown,
averaged over volume, Ly$\alpha$ emissivity is relatively
insensitive to the luminosity of the central object. This results from
the gas temperature having only a weak dependence on luminosity for
the relevant parameter range and from hydrogen being fully
ionized. Emissivity is suppressed on small scales near very luminous
sources due to suppression of the recombination rate at high ($\gg
10^5$\,K) temperatures.

We have also studied the emissivity of other lines and, in particular,
that of [\ion{O}{3}] $\lambda 5007$ forbidden line
(Fig. \ref{lumi_lumi}). The line emissivity in this case is very
different from that of Ly$\alpha$ and other hydrogen-like lines
since the recombination rate is much lower (\ion{O}{3}\ and
\ion{O}{4}\ abundances in the gas are negligible). Relatively neutral
gas, which is required for efficient [\ion{O}{3}] $\lambda 5007$
emission, becomes abundant only on large scales and even then only for
relatively low luminosity quasars. As quasars get brighter the peak
\ion{O}{3}\ emissivity is shifted to larger scales where cooler gas is found. We emphasize that,
unlike Ly$\alpha$ emission, [\ion{O}{3}] $\lambda 5007$ emission is
sensitive to gas composition and is roughly linear with it.

Our calculations allow us to predict how quasar halos would appear for
different emission lines. We show an example for a Ly$\alpha$ image in
Fig. \ref{lumi_lumi} for the case of $\theta=70^\circ,~f_M=5,~L_{46}=1$ and
assume that our line-of-sight is along the symmetry axis of the
quasar. The image is symmetric with the surface brightness profile
being roughly $\propto b^{-1}$ from the center (neglecting the effects
of scattering; see below).  We note that asymmetric images are
obtained if the axis of symmetry is at an angle to our line-of-sight,
as likely to be the case in general.

Different lines have different emissivity profiles and hence different
luminosities and inferred nebular sizes. Table 2 lists a few of the
more important lines which could, in principal, be observed in quasar
haloes. Ly$\alpha$, \ion{C}{4} $\lambda 1548$, and \ion{O}{6}
$\lambda 1035$ are the strongest transitions with the latter two
lines depending roughly linearly on the Metallicity. The typical
nebular extent (from which most of the flux is emitted) of low ionization 
lines is  $>100$\,kpc. This is in contrast to the apparent size of high ionization lines 
which come from compact, highly ionized regions of the halo (e.g., the nebular extent 
of  \ion{O}{8} X-ray lines is $\gtrsim 10$\,kpc for  $L_{46}=1$).

\subsubsection{Scattering}

In addition to line emission there is scattering (i.e. absorption
followed by emission) of the photons emitted by the quasar. Here broad
emission line photons can be
scattered off gas clouds in the halo provided the neutral hydrogen
fraction is large enough.  Owing to the low gas density, the
population of excited levels is negligible and so only resonance lines
are able to scatter efficiently. To correctly calculate the surface
brightness profile due to photon scattering one requires detailed
knowledge of the 3D distribution of clouds in real and velocity
space. In addition, the fraction of photons reaching the halo depends
on the amount of intrinsic absorption and the relative velocities of
the emission line and halo gas. Detailed calculations should include
the effect of multiple scatterings even in this relatively optically
thin medium. 

Despite the overall complexity of scattering problems, it is
relatively straightforward to estimate the maximum scattered flux
expected for different lines. This is simply the flux in the broad
emission line with velocity dispersion, $\sigma_l$ which is
intercepted by halo clouds with velocity dispersion, $\sigma_c$ given
by roughly $\sigma_c/\sigma_l\simeq 10\%$ (here we used $\sigma_l
\simeq 2000$; e.g., Vanden Berk et al. 2001). Given our quasar SED and
taking the mean rest equivalent width for Ly$\alpha$ to be $\sim
90$\AA\ (e.g., Zheng et al. 1997; Vanden Berk et al. 2001) we find
that the the maximum scattered Ly$\alpha$ luminosity, $L_{\rm
Ly\,\alpha}^{\rm scat,max}\sim 2\times 10^{43}\left [ 1-{\rm sin} ( \theta )
\right ] L_{46}~{\rm erg~s^{-1}}$.  Table 2 lists the results for
other lines based on the same line-of-reasoning and shows that the contribution 
of scattered broad line region photons to the large scale nebular emission is 
probably less than 50\% for all prominent lines (and for $L_{46}<10^2$).

\begin{table}
\begin{center}
{\sc TABLE 2 \\ Emission Line Properties of Quasar Nebulae}
\vskip 4pt
\begin{tabular}{llllllll}
\tableline
\tableline
Line ID & $L $ & Size & $\frac{L^{\rm scat,max}}{L+L^{\rm scat,max}}$\\
 & $[{\rm erg~s^{-1}}]$ & $[{\rm kpc}]$ &  [ ] \\
\tableline
\ion{H}{1}\ Ly$\alpha$ & $3\times 10^{43}$ & 150 & $0.25$\\
\ion{O}{6}\ $\lambda 1035$ & $4\times10^{42}$ & 80 & $0.25$ \\
\ion{N}{5}\ $\lambda 1238$ & $2\times10^{42}$ & 150  & $0.25$ \\
\ion{C}{4}\ $\lambda 1548$ & $2\times10^{42}$ & 200 & $0.5$ \\
\ion{H}{1}\ H$\alpha$ & $1\times 10^{42}$ & 100 & - \\
\ion{H}{1}\ H$\beta$ & $8\times 10^{42}$ & 150 & - \\
\ion{C}{3}\ $\lambda 977$ & $7\times10^{41}$ & 200 & - \\
\ion{O}{8}\ $\lambda 18.97$ & $4\times 10^{41}$ & 20 & ?\\
$[$\ion{O}{3}$]$\, $\lambda 5007$ & $2\times10^{41}$ & 200 & - \\
\tableline
\end{tabular}
\vskip 2pt
\parbox{4.2in}{ 
\small\baselineskip 9pt \footnotesize \indent Luminosities for
doublets are the sum for individual transitions. Size is defined \\ as
the extent of the emitting nebula which contains 90\% of the flux for a given \\
transition. The last column is an order of magnitude estimate for
the maximum \\ luminosity in scattered light (see text). Transitions
that do not scatter are marked \\ with ``-''  while those for which the
broad emission line flux is very uncertain \\ are marked with ``?''
(see text). $L_{46}=1,~\theta=70^\circ$ and $f_M=5$ are assumed.}
\end{center}
\end{table}

To better evaluate the flux due to broad emission line photons
scattering off halo gas, we define an effective optical depth over the
(dynamically broadened) absorption line profile of halo gas,
$\tau_{\rm eff}(r) = C(r)\tau(r)$ where $C(r)$ is the geometric
covering factor of clouds at a given distance (i.e., the number of
clouds intercepted at some distance interval) and $\tau(r)$ is the
line optical depth of a single cloud at position $r$. In our model,
$C(r)$ is large on small scales due to the clouds density peaking
toward the center yet, $\tau(r)$, is low since the gas is more
ionized. For Ly$\alpha$, $\tau_{\rm eff}$ is decreasing with
decreasing $r$. On scales larger than $\sim 30$\,kpc, clouds become
opaque to Ly$\alpha$ and $\tau_{\rm eff}\propto C\propto r^{-2}$. We
have approximated the problem by running monte-carlo simulations
(e.g., Zheng \& Miralda-Escude 2002; Z. Zheng, private communication)
of photons emitted by a point source and being scattered off halo gas
with $\tau_{\rm eff}$ being of order a few and having the radial
dependence described above. We find that the surface brightness of the
scattered photons is $\propto b^{-2.5}$ and therefore that most of the
scattering occurs in the inner nebular regions (within $\sim
30$\,kpc). The results are shown in the inset of Fig. \ref{lumi_lumi} for the case of
$L_{46}=1$. We note, however, that these provide only 
order-of-magnitude estimates since the gas kinematics in the central
regions of the halo is rather uncertain.

\subsubsection{Detectability}

It is interesting to examine the detectability of line emitting
nebulae around quasars within the framework of our model. Here we
focus on Ly$\alpha$ nebulosity and study its observed luminosity and
size as a function of the limiting flux of the observation (see
Fig. \ref{plot_sensitivity}). As before, $f_M=10,~\theta=70^\circ$ and
$L_{46}=10$ are assumed. We also examine these quantities for several
redshifts assuming the observed nebulae at those redshifts are similar
so that the surface brightness is $\propto (1+z)^4$. Our calculations
indicate that the limiting flux required to observe extended
Ly$\alpha$ emission around $z=1$ quasars is better than $\sim
10^{-16}~{\rm erg~s^{-1}~cm^{-2}~asec^{-2}}$.  In particular, our
model suggests that the observed nebular extent is sensitive to the
limiting flux with a factor two increase in sensitivity allowing to
detect emission from an order of magnitude larger volume. At low-$z$,
a limiting flux of a few$\times 10^{-18}~{\rm
erg~s^{-1}~cm^{-2}~asec^{-2}}$ is required to probe the full extent of
the nebulosity and correctly estimate its luminosity. Once the
appropriate sensitivity is reached, our model suggests that large
scale emission line nebulae in general, and Ly$\alpha$ nebulae in
particular, should be a rather common phenomenon around quasars. More
massive halos or quasars having a wider ionization cones are better
emitters and their halos can be observed to higher-$z$. Our
calculations suggest that Ly$\alpha$ scattering can have a substantial
contribution to the observed luminosity (compare models B,C and D in
figure \ref{plot_sensitivity}). As expected, halos with lower mass of
cool gas ($f_M=1$) are poor emitters (compare models A and B in figure
\ref{plot_sensitivity}). Similar considerations also apply for metal
lines whose detection can be used to put more stringent constraints on
the ionization level of the emitting gas as well as estimate its mean
metallicity. For example, the \ion{N}{5}\,$\lambda 1238$ emission line
could have a flux comparable  to that of Ly$\alpha$ for $\gtrsim$solar
metallicity gas (see table 2).\\

\subsubsection{Comparison with observations}

\begin{figure}
\includegraphics[width=\hsize]{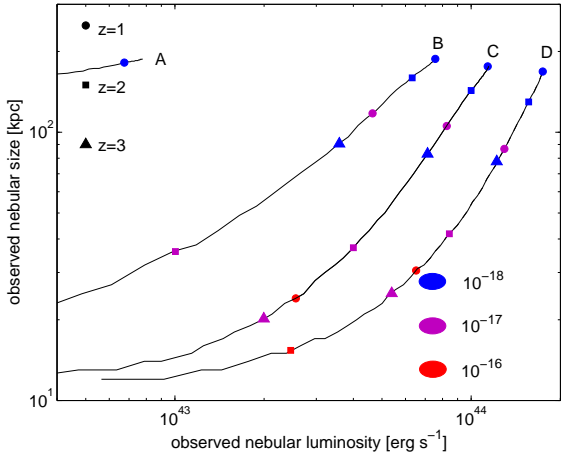}
\caption{The observed Ly$\alpha$ nebula size (containing 90\% of the total obsrved flux) vs. the observed luminosity for a halo illuminated by a  $L_{46}=10$ quasar with an opening angle $\theta=70^\circ$. The following models are considered: (A) $f_M=1$ with no line scattering. (B) $f_M=10$ with no line scattering. (C) $f_M=10$ with scattered Ly$\alpha$ luminosity, $L_{\rm Ly\alpha}^{\rm scat}=0.5L_{\rm Ly\alpha}^{\rm scat, max}$. (D) $f_M=10$ and $L_{\rm Ly\alpha}^{\rm scat}=L_{\rm Ly\alpha}^{\rm scat, max}$. We assume no evolution in the nebular properties with redshift. Placing the same nebula at different redshifts (denoted by different symbols; see legend) decreases its observed luminosity and size for a survey with a given limiting flux (denoted by colors in units of ${\rm erg~s^{-1} cm^{-2} asec^{-2}}$).  Clearly, unless the nebula is close and/or the sensitivity is high, only the brighter inner part of the nebula is observed which results in an under-estimation of its luminosity and size (see text). In particular, sensitivities of order $10^{-18}~{\rm erg~s^{-1} cm^{-2} asec^{-2}}$ are required to probe the full extent of quasar nebulae at $z\lesssim 2$ (blue square and circle). }
\label{plot_sensitivity}
\end{figure}

Extended narrow-line emission around quasars and radio galaxies has
been observed for more than two decades and the accumulation of data
has shown that giant Ly$\alpha$ nebulae can be found around QSOs, up
to scales of $\sim$ 100 kpc (e.g., Wampler et al. 1975, Stockton 1976,
Bremer et al. 1992, Boroson, Oke, \& Green 1982, Bergeron et al. 1983,
Boroson, Persson, \& Oke 1985, Stockton \& MacKenty 1987, Reuland et
al. 2003, Labiano et al. 2005).  Early studies focused mostly on radio
loud quasars but more recent ones showed that radio-quiet quasars also
show extended Ly$\alpha$ emission on large scales.
It is interesting to note that Hu et al. (1991) observed a sample of 7
radio-quiet quasars down to a limiting flux of $2\times10^{-16}
\rm{erg s}^{-1} \rm{cm}^{-2}$ and found no extended Ly$\alpha$
emission. However, subsequent and more sensitive observations showed
that quasar nebulae are commonly observed around radio-quiet quasars
(Weidinger et al. 2005, Bremer et al. 1992., Steidel et al. 1991, Hu
et al. 1996, Petitjean et al. 1996).  These authors report Ly$\alpha$
luminosities in the range $7-58\times10^{42}\,\rm{erg s}^{-1}$, values
which are largely consistent with our predictions as shown in Fig. 12.
Christensen et al. (2006) report Ly$\alpha$ luminosities from $z\sim
3$ nebulae which are of order $\sim 10^{43}~{\rm erg~s^{-1}}$ and
extend up to a median value of 15\,kpc (with some extending up to
60\,kpc) for a flux limit of $\sim 10^{-17}~{\rm erg~
cm^{-2}~s^{-1}~asec^{-2}}$. This is consistent with quasar halos having a considerable 
amount of gas in cool form (model A in figure \ref{plot_sensitivity}
corresponding to $f_M=1$ with no line scattering is excluded by the
data; c.f. Binette et al. 2006).  Christensen et al. place limits on
\ion{C}{4}\ line emission down to fluxes $\sim 10$ times lower than
Ly$\alpha$. In our model this is roughly the predicted flux ratio of
the lines for 0.1 solar metallicity.  More detailed modeling is,
however, beyond the scope of this paper.

The statistical properties of quasar halos and their underlying
physics are poorly known and few models have been devised to account
for them (e.g., Heckman et al. 1991). Common explanations for the
existence of such line emission nebulae include gas accretion onto the
central object (Haiman \& Rees 2001) and matter ejection by the quasar
and its host galaxy (e.g., Heckman et al. 1991). Moreover, the
interpretation of the data is controversial: while several teams have
reported high pressures of order $P\sim10^{6-7} \rm{cm}^{-3}$\,K for
radio loud quasars (RLQ), a study by Fu \& Stockton (2006) points to
lower pressures, of order $10^4\rm{cm}^{-3}$ K. These estimates come
from compact emission regions of [O\,II] and [O\,III] lines and from
absorption by excited states (e.g., Hamann et al. 2001). In our model
for the halo of an $L^\star$ galaxy the pressure is even lower,
$P\sim$a few$\times 10^2$ cm$^{-3}$.  It is possible that RLQ are
different from radio-quiet quasars (RQQ) given their overall higher
Ly$\alpha$ luminosities (e.g., Christensen et al. 2006) and their
apparently large X-ray and [OIII] emission on small scales (e.g.,
Crawford \& Fabian 1989). Our model predicts relatively faint and
diffuse [\ion{O}{3}] emission from the halo which seems to be at odds
with recent detections of [\ion{O}{2}] and [\ion{O}{3}] emitting
filaments around RLQs. Deep optical observations of low-$z$ RQQs may
help to determine whether there is a relatively dense and filamentary
structure in quasar halos in addition to the more diffuse component
considered here.

\section{Discussion}

\subsection{Galaxies}

Our model implies considerable mass in cool gas out to the virial
radius of $L^\star$ galaxy halos at $z\sim 1$. As explained above,
this assumes a unit covering factor for strong absorption within
50\,kpc of the galaxy. The lower covering factors suggested by
Bechtold \& Ellingson (1992) and Tripp \& Bowen (2005) would actually
lead to an agreement between our model predictions and the results of Maller
\& Bullock (2004) who estimated the mass of cool gas in the Galaxy
halo to be $2\times 10^{10}~{\rm M_\odot}$ and argued for consistency
with observations of high-velocity clouds.  At present, radio surveys
do not possess the required sensitivity to probe low column density,
$\sim 100$\,kpc diffuse gas in 21\,cm emission, as predicted by our
model.  We caution, however, that it is not clear whether our model
predictions for $z\sim 1$ objects are directly applicable to present
day galaxies. In particular, Nestor et al. (2005) find evidence for
evolution of cool gas properties with cosmic time whereby the
occurrance of strong systems appreciably declines from $z=1$ to
$z=0.5$. Understanding the distribution of cool gas around present day
galaxies will require extensive UV spectroscopic surveys.

The notion of massive galaxy halos has been promoted by Fukugita \&
Peebles (2006) to provide a solution to the missing baryon problem. In
their model, the gas mass up to the virial radius is a few$\times
10^{11}~{\rm M_\odot}$. Their total gas mass estimates provide a
natural explanation to the missing baryon problem and are consistent
with a large range of observational constraints which are independent
of those considered here.  Their model did not consider the possible
existence of cool gas beyond the extent of the disk. Here we have
shown that converting a fraction of the hot plasma into cool clumps
allows us to reproduce a number of observational constraints from
absorption line studies, while not significantly changing the total
amount of gas within the virial radius (see Fig. 5).

More work is needed to establish how much mass there is in the halos
of present day galaxies. On the observational side, better estimates
of the covering fraction of cool gas around galaxies are required,
possibly as a function of galaxy type.  On the theoretical side,
numerical simulations of galaxy formation need to include realistic
cooling rates and reach high resolutions to reliably trace cool and
condensed forms of gas. Understanding the properties of the dilute and
ellusive, yet likely very massive, cool gas component in galaxy halos
is essential for understanding galaxy formation and evolution from the
early Universe, to present times and beyond.

\subsection{Quasars}

Quasars are known to reside in galaxies yet their large scale
environments are poorly understood. The recent discovery of
\ion{Mg}{2}\ absorbing gas around those objects (Bowen et al. 2006,
2007) suggests an interesting link between galaxy and quasar
environments. This implies that the presence of a quasar can be used to shed
light on the distribution of matter in the halos of galaxies.

We have shown that it is possible to construct a model which unifies
several distinct astrophysical phenomena pertaining to intervening
absorption, associated absorption, and quasar emission nebulae. In our
model, these phenomena are different manifestations of the same
physical entity, that being of cool gas condensations distributed
within the halo of galaxies.  If our model is correct then the
properties of cool gas in the vicinity of galaxies may be revealed by
studying the influence of an active nucleus on its thermal and ionization
structure. In particular, the study of associated absorption may help
to constrain the density and distance of halo gas from the
quasar. Cool gas around galaxies is notoriously difficult to detect in
emission.  Nevertheless, the presence of an ionizing source enhances
emission and renders the detection of circum-galactic material more
accessible. Furthermore, by studying the nebulosities around quasars,
one can obtain a 2D-spatial$\times$1D-velocity picture of the
gas. Mean quantities pertaining to the mass of the gas and its
metallicity can be more easily deduced for individual objects.

In addition to shedding light on gas distribution in the halos of
galaxies, it is possible to deduce important conclusions about quasar
physics: the recent results by Bowen et al. (2006, 2007) indicate a covering
factor of order unity for cool gas in the transverse direction around
quasars in contrast to the much lower covering factor for associated
\ion{Mg}{2}\ absorption. Our
model implies that this is merely a consequence of the quasar
unification scheme: gas in the radial direction to the quasar is
heated to high temperatures while that which is at right angles to the
quasar ionization cone remains unaffected.  This is not a trivial
result since, although the unification scheme is well established for
active galactic nuclei, it is less secure for bright quasars at
high-$z$. Further tests of this conjecture may be carried out by
comparing model predictions to the occurrence of associated absorbers
for a wide range of ionization levels as well as by studying emission
line nebulae in detail.

Our model can reproduce the emission and absorption properties of
quasars by requiring quasar halos to be more extended and therefore
more massive than those of $L^\star$ galaxies (c.f. Serber et
al. 2006).  Such a scaling combined with the radiation properties of
quasars can simultaneously explain the anisotropic absorption
properties and the Ly$\alpha$ nebulae observed on scales reaching
$\sim$100 kpc. Finally, if the correlation found between the size of
cool gaseous halos and galaxy luminosity (Steidel et al. 1997,
Guillemin \& Bergeron 1997) can be extended to quasars then this
implies that quasar hosts are a few$\times L^\star$, which is in
agreement with results from Jahnke et al. (2004).

Better
understanding of quasar emission line nebulae may allow us to deduce
the long-term light-curve behavior of quasars. This will require
better understanding of the matter distribution in quasar halos which
can be achieved by deep imaging and spatially resolved spectroscopy of
quasar environments. The detection of orphan nebulae (where the active nucleus is obscure) may provide a new
means for identifying type-II quasars. Combining large samples of
emitting and absorbing gaseous nebulae may allow us to statistically
constrain the mass of cool gas in quasar halos as well as the opening
angle of quasars. This has implications for re-ionization and
background radiation determination.  A survey of emission line nebulae
around a large number of quasars may also reveal environmental
differences among quasar types (e.g., the RQQ-RLQ dichotomy).

The detection of metal emission lines, whose presence (depending on
the gas composition) is predicted by our model, may allow to estimate
the metallicity of gas in the halo and test models for metal enrichment
of galaxies and the inter-galactic/intra-cluster medium.

\section{Summary}

We have presented a phenomenological model for the distribution of
cool gas around $L^\star$ galaxies, calibrated with a wide range of
observational constraints from absorption line studies (rest
equivalent width distribution of \ion{Mg}{2}\ absorbers, HI column
density, etc.). We argue that the halos of $L^\star$ galaxies are
filled with gaseous clouds with sizes of order 1\,kpc, masses of
$\sim10^6 M_\sun$ and particle densities $\sim 10^{-2}\rm{cm^{-3}}$.
The total amount of cool gas within the virial radius of $L^\star$
galaxies depends on the covering factor for strong \ion{Mg}{2}\
absorption (which is a matter of debate) and is likely to be in the
range $10^{10}<M_{\rm cool}<10^{11}\,M_\sun$, within the virial
radius.

By assuming self-similarity, we show that the obtained solution for
the distribution of cool gas around galaxies, if appropriately scaled,
can reproduce the properties of cool gas seen quasars. Our model
simultaneously provides an explanation for the L$\alpha$ nebulosities
observed observed on $\sim100$ kpc scales around quasars as well
as the main properties of the cool gas seen in absorption around
quasars.

Comparison of model predictions with future surveys will shed light
on the missing baryon problem and will deepen our understanding with
respect to galaxy formation and quasar activation.

\acknowledgements

We thank Gary Ferland for creating and maintaining {\sc cloudy} as a
publicly available code. We are grateful to Z. Zheng for invaluable help with
the Ly$\alpha$ scattering calculations. We thank K. Jahnke, N. Murray,  and 
D. York for commenting on an earlier version of this paper. This
research has been supported by NASA through a Chandra Postdoctoral
Fellowship award PF4-50033. DVB is funded through NASA Long Term Space
Astrophysics Grant NNG05GE26G.

\appendix

Here we describe the time-dependent ionization and thermal structure
of an initially cool cloud which is exposed to a quasar's ionizing
radiation.  As the quasar phenomenon is short lived compared to the
Hubble time (the quasar lifetime, $\tau_q\sim 10^7$\,years; e.g., Worseck \& Wisotzki
2006), it is instructive to first consider a few relevant timescales
of the problem:
\begin{itemize}
\item
The dynamical timescale of a halo of size $R_{\rm halo}$ at a virial temperature,
$T_v$ (corresponding to some sound speed, $v_s$) is
\begin{equation}
\tau_{\rm dyn}\sim \frac{R_{\rm halo}}{v_s} \simeq 10^9
\frac{r}{100\,{\rm kpc}} \left ( \frac{T_v} {10^7{\rm K}} \right )
^{-1/2}~{\rm yr}
\label{tdyn}
\end{equation}
\item
The sound crossing timescale for a cloud of size $2R_{\rm cloud}$ at
temperature $T$ is
\begin{equation}
\tau_{\rm cross} \sim \frac{2R_{\rm cloud}}{v_s} \simeq 10^8 \frac{R_{\rm cloud}}{1\,{\rm kpc}} \left ( \frac{T} {10^4{\rm K}} \right ) ^{-1/2}~{\rm yr}
\end{equation}
\item
The photoionization timescale for ion $X$ is
\begin{equation}
\tau_{\rm ion}^{X}=\left ( \int \frac{ \sigma_X(E) L_E}{4\pi r^2E}
dE \right )^{-1}\sim 10^3 L_{46}^{-1} \left ( \frac{r}{{\rm 10^2kpc}} \right )^2\,{\rm yr}
\end{equation}
where $\sigma_X$ is the cross-section, $E$ the energy, $L_E$ the flux per unit energy,
$r$ the distance from the quasar.  We assumed a mean ionizing photon
energy of $E=10$\,eV and an ionization cross-section of $10^{-19}~{\rm
cm^{2}}$ (roughly that of \ion{Mg}{2}).
\item
The recombination timescale for ion $X$ is
\begin{equation}
\tau_{\rm rec}^{X}=\frac{1}{ n \alpha_X(T)} \sim
5\times 10^4\left ( \frac{n}{0.02{\rm cm^{-3}}} \right ) ^{-1}~{\rm yr}
\end{equation}
where $n$ is the number density of atoms and $\alpha_X$ the
recombination coefficient (taking only the radiative term for
\ion{Mg}{2}).
\end{itemize}
The photoionization and recombination timescales are the
shortest in the problem. The cloud and halo dynamical timescales
(approximated by the sound-crossing time) are probably longer than the
quasar lifetime, $\tau_q$; hence, while the ionization structure of
the clouds can vary rapidly (see below), the structure of the halo and
embedded clouds remains approximately constant during the quasar
lifetime. We note, however, that the  sound-crossing timescale for
the smallest clouds in our model could be as short as $\sim
10^7$\,years if their temperature rises to high values once the quasar
activates. We choose to ignore this complication here since (a) the
quasar lifetime is poorly known and (b) the evaporation timescale is a
few times the sound-crossing timescale (e.g., Bertoldi 1989). Also,
due to the spectrum of the cloud's size-distribution, the evaporation
of the smallest clouds will make little difference to our final
conclusions.

\begin{figure*}
\plottwo{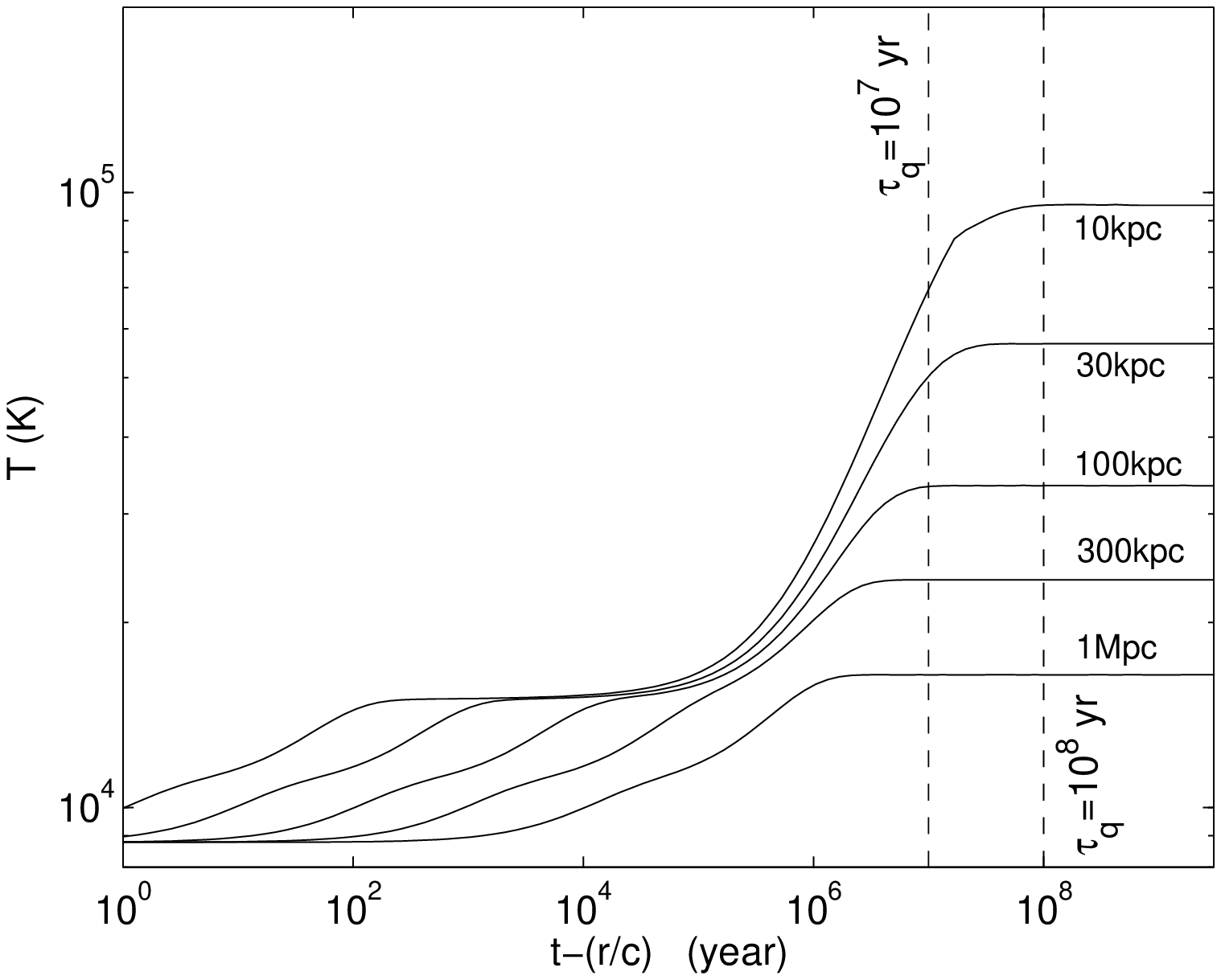}{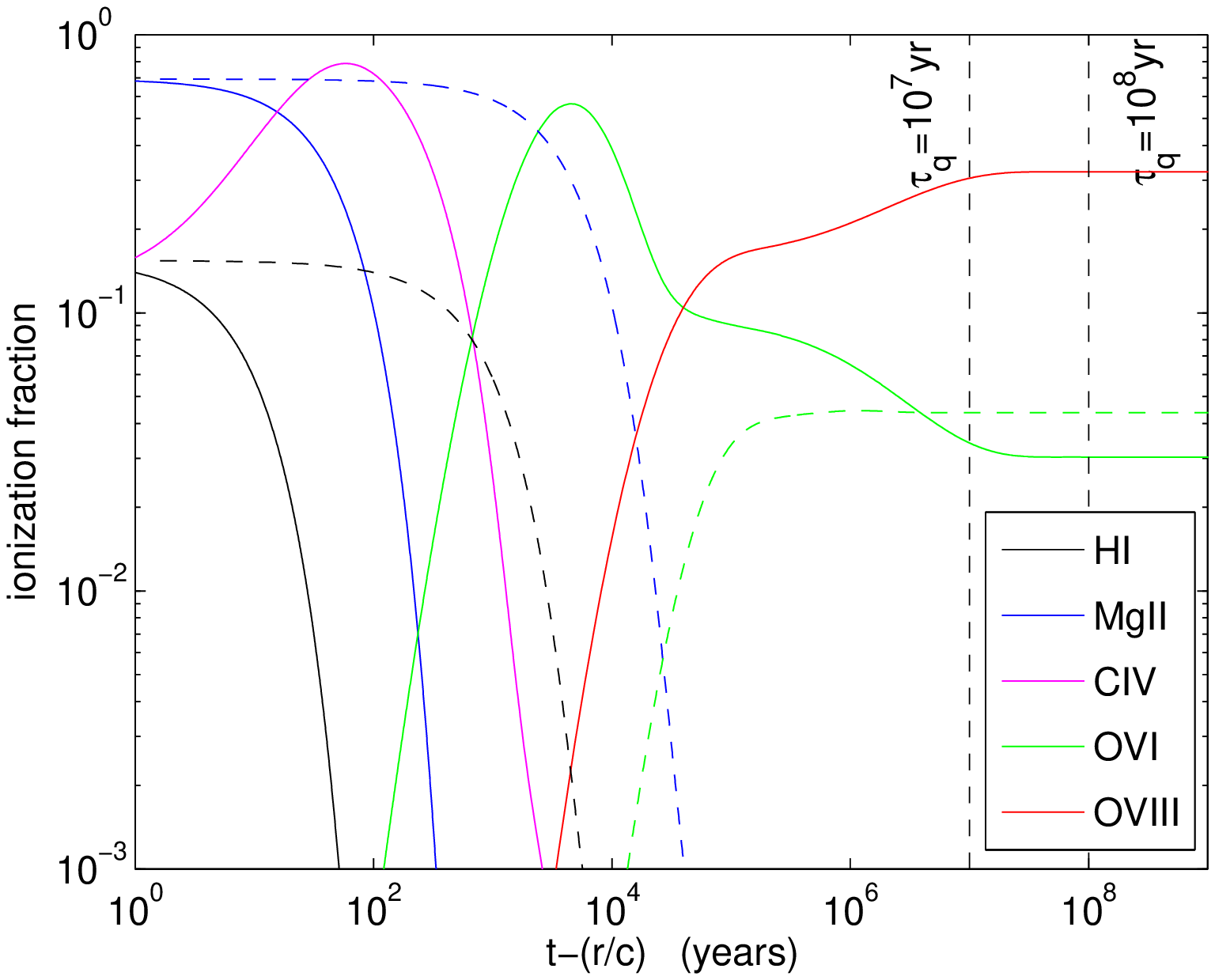}
\caption{Time dependent thermal structure of the halo after a quasar
ignites at $t=0$ (and a cloud at a distance $r$ from it is then exposed at time $t+r/c$) assuming $L_{46}(t)=\Theta(t)$ where $\Theta(t)$ is a
step function. Also plotted are estimates for typical quasar
lifetimes. Time dependent photoionization effects are
important and prevent the gas from reaching very high temperatures
(close to the Compton temperature) over the quasar
lifetime. \emph{Right:} Ionization fractions for several abundant
elements as a function of time. Colors denote different ions (see
legend) while different lines denote gas at different distances from
the quasars: 30\,kpc (solid), and 300\,kpc (dashed) lines. The rapid
ionization of all low level ions is clearly seen and takes place over
$\lesssim 10^4$\,years. Such gas will therefore disappear from the
line-of-sight and would become transparent. We note the general
agreement between the ionization timescales obtained and the analytic
approximation (equation A3).}
\label{orly}
\end{figure*}

In addition to the above analytic approximations, we have numerically
computed several time-dependent photoionization and thermal
models. This is necessary to quantify just how important the effects
of the varying quasar flux are compared to time independent
calculations and to see which ions are affected most. The algorithm used for
these calculations is fully described in Gnat \&
Sternberg (2006). 

At the beginning of the calculation we assume that the cloud is exposed
to the meta-galactic field and is in ionization and thermal
equilibrium. We then turn on a quasar and assume a step function
description for its light curve, i.e., the quasar if off for $t<0$ and
on at $t\geq 0$. The quasar is assumed to maintain a constant flux
level to $t\rightarrow \infty$. The calculations are carried out under
isochoric conditions, as appropriate given the above considerations.

The thermal calculations are shown in Fig. \ref{orly} for the case
of $L_{46}=1$, and demonstrate the limitations of the assumption of
photoionization and thermal equilibrium. The gas heats up
very quickly near the quasar where the ionizing flux is high and less
so on larger scales. Typically, the gas within the inner 50\,kpc heats
up to $\sim 2\times 10^4$\,K over a period of $\sim 10^4$ years
(cf. equations A3,A4). Further heating occurs on somewhat longer
timescales which can be comparable to the quasar lifetime. Unless one
happens to intercept young quasars then the inner $50$\,kpc region of
the halo would be heated to $\lesssim 10^5$\,K over the quasar
lifetime. We conclude that, for moderate luminosity quasars, the gas
in the central regions of the haloes can be efficiently heated to
$\lesssim 10^5$\,K. Nevertheless, further heating is delayed and may
take longer than the quasar duty-cycle or other relevant dynamical
timescales of the problem.  Assuming $\tau_q$ is similar for all
quasars then higher temperatures can be reached in the inner regions
of the haloes of bright quasars since the ionization timescale is
shorter and the flux higher.

We have also calculated the time-dependent ionization structure of the
halo and trace several of the ionization levels in Fig.
\ref{orly}. As shown, all the low ionization stages usually probed
by absorption line studies quickly attain negligible abundances and are not
likely to be observed unless the quasar brightened over short
timescales.  For example, \ion{Mg}{2}\ and \ion{H}{1}\ would be
immediately ionized and the gas is quickly heated. Time-dependent
ionization deviations from a steady-state solution can be important
but mainly for the high ionization levels whose absorption lines would
appear in the X-ray region of the quasar spectrum.

The above analysis suggests that it may be possible to treat the halos of
quasars as having the same structure as galaxy halos since little
structural changes are likely to occur over the quasar lifetime by
radiation effects alone. We note that this is may not be the case if
quasar activity is a recurring phenomenon or if they have 
additional means of affecting their environment (e.g., via jets, winds,
etc.). To conclude, for the purpose of this work, steady-state photoionization
calculations can be adequately used to describe the ionization and
thermal state of the halo gas at all times.

\end{document}